\documentclass[11pt,a4paper]{article}
\usepackage[utf8]{inputenc}
\usepackage[english]{babel}
\usepackage{indentfirst}
\usepackage{cite}
\usepackage{cmap}
\usepackage{longtable,array}
\usepackage{caption}
\usepackage{bm}
\usepackage{rotating}
\usepackage{graphicx}
\usepackage{amsmath,amssymb,amsthm,amsfonts}
\usepackage{mathrsfs}
\usepackage[mathscr]{eucal}
\usepackage{hyperref}
\hypersetup{colorlinks=true,linkcolor=blue,citecolor=blue,urlcolor=blue,breaklinks=true}
\newcommand{\doititle}[2]{\href{https://doi.org/#1}{#2}}

\allowdisplaybreaks
\newtheorem{theorem}{Theorem}

\newtheorem{corollary}[theorem]{Corollary}
{\theoremstyle{definition}
\newtheorem{remark}{Remark}
\newtheorem{definition}{Definition}
\newtheorem*{remark*}{Remark}
\newtheorem*{definition*}{Definition}
}
\date{}

\newcommand{\p}{\partial}
\newcommand{\tr}{\mathop{\rm tr}\nolimits}
\newcommand{\const}{{\rm const}}
\newcommand{\lsemioplus}{\mathbin{\mbox{$\lefteqn{\hspace{.77ex}\rule{.4pt}{1.2ex}}{\in}$}}}
\newcommand{\rsemioplus}{\mathbin{\mbox{$\lefteqn{\hspace{.76ex}\rule{.4pt}{1.2ex}}{\ni}$}}}
\newcommand{\todo}[1][\null]{\ensuremath{\clubsuit}}
\newcommand{\noprint}[1]{}

\begin{document}

\noindent
{\LARGE\bfseries Group classification of a class of systems\\ of nonlinear diffusion equations. I\par}

\vspace{4mm}
\noindent
{\large Oleksandr~A.~Voloshyn$^{\dagger,*}$, Mykola~I.~Serov$^\ddagger$ and Olena~O.~Vaneeva$^\dagger$\par}

\vspace{4mm}
\noindent{\itshape\small
$^\dagger$Institute of Mathematics, National Academy of Sciences of Ukraine, Kyiv, Ukraine\\[1mm]
$^\ddagger$V.~G.~Korolenko Poltava National Pedagogical University, Poltava, Ukraine
}\par

\vspace{3mm}
\noindent E-mails: voloshyn@imath.kiev.ua, mserov4@gmail.com, vaneeva@imath.kiev.ua

\begingroup\renewcommand{\thefootnote}{\fnsymbol{footnote}}\footnotetext[1]{Corresponding author.}\endgroup

\vspace{5mm}\par\noindent\hspace*{5mm}\parbox{150mm}{\small
We study the Lie symmetries of a class of systems of (1 + 1)-dimensional nonlinear diffusion equations. The equivalence algebra and the kernel of the maximal Lie invariance algebras of this class are found. We prove that there exist exactly nine inequivalent subclasses of this class for which the corresponding diffusion systems admit Lie symmetry extensions. We also carry out the exhaustive group classification of the systems of nonlinear diffusion equations from the first subclass.
}\par\vspace{4mm}


\vspace{5mm}

\section{Introduction}

In the general case, a system of nonlinear diffusion equations for a two-dimensional vector function $U\in\mathbb{R}^2$ has the form
\begin{gather}\label{eq:mainsystem}
	U_t=\p_x \left[F(U)U_x\right].
\end{gather}
Here and in what follows,
\begin{gather*}
		U:=\begin{pmatrix}
			u^1 \\
			u^2
		\end{pmatrix}
		\in \mathbb{R}^2, \quad F(U):=\begin{pmatrix}
		f^{11} & f^{12}\\
		f^{21} & f^{22}
	\end{pmatrix}, \quad U_t=\dfrac{\p U}{\p t}, \quad	U_x=\dfrac{\p U}{\p x}, \quad	\p_t=\dfrac{\p}{\p t}, \quad	\p_x=\dfrac{\p}{\p x},
	\end{gather*}
$u^a=u^a(t,x)$ are functions describing the concentrations or densities of two interacting substances, $t$ and $x$ are the time and space variables, respectively, $f^{ab}=f^{ab}(u^1,u^2)$ are diffusion coefficients, which are arbitrary smooth functions of their arguments, $F(U)$ is a nonzero matrix, the indices $a,b,c,d$ run through $\{1,2\}$, and summation from 1 to 2 is understood over repeated indices.
	
The diffusion system~\eqref{eq:mainsystem} is used in various areas of science and technology to model and understand diffusion processes. It describes heat conduction in solids, mass transfer in liquids and gases, diffusion in biological systems, chemical reactions in solids, distribution of concentrations in industrial processes, migration of pollutants in soil, temperature distribution in the ocean, ion diffusion in biochemistry, diffusion in electronic devices, and dispersion of odors in the atmosphere, among many other phenomena.

Since a system of the form~\eqref{eq:mainsystem} contains four arbitrary elements $f^{ab}=f^{ab}(u^1,u^2)$, it defines a class of systems of (1+1)-dimensional nonlinear differential equations of second order.

Our aim is to study class~\eqref{eq:mainsystem} from the point of view of symmetry analysis of differential equations, that is, to describe the symmetry properties of system~\eqref{eq:mainsystem} depending on the diffusion coefficients $f^{ab}=f^{ab}(u^1,u^2)$. This is called the problem of group classification of class~\eqref{eq:mainsystem}.

The group classification problem for partial differential equations (PDEs) was initiated by L.~V.~Ovsyannikov in~\cite{ovs-gf} for the class of scalar diffusion equations
\begin{gather}\label{eq:nonlinear-ovs}
	u_t=\p_x \left(f(u)u_x\right),
\end{gather}
where $u=u(t,x)\in\mathbb{R}^1$, $f=f(u)$ is an arbitrary nonzero smooth function, and $f_u\ne0$.

Since then, numerous papers have been devoted to the symmetry properties of nonlinear differential equations; see, for example, the collection of results in~\cite{Ibragimov1994V1}. We mention some works most closely related to our study.

Invariant solutions of the nonlinear heat-conduction equation with a source were studied in~\cite{dor-kniaz-svir}:
\begin{gather*}
	u_{t}=\p_{x}(D(u)u_{x}) + g(u),
\end{gather*}
where $u=u(t,x)\in\mathbb{R}^1$ and $D=D(u)$ and $g=g(u)$ are functions of $u$.
The symmetry properties and exact solutions of the class of equations
\begin{gather*}
u_t =h(u)u_{xx}
\end{gather*}
with an arbitrary nonzero smooth function $h=h(u)$ and $u=u(t,x)\in\mathbb{R}^1$
were studied by V.~A.~Tychynin in~\cite{tychynin}.
The symmetry properties of the class of convection--diffusion equations
\begin{gather*}
	u_{t}=\p_{x}(D(u)u_{x}) + f(u)u_{x},
\end{gather*}
where $u=u(t,x)\in\mathbb{R}^1$ and $D=D(u)$ and $f=f(u)$ are arbitrary nonzero smooth functions,
were investigated in~\cite{oron-rose,yung-verb-baveye,edwards}.
The group classification of the class of filtration equations
\begin{gather*}
	u_{t}=G(u_{x})u_{xx}
\end{gather*}
with an arbitrary nonzero smooth function $G=G(u_x)$ and $u=u(t,x)\in\mathbb{R}^1$ was performed in~\cite{ahat-gaz-ibr}.
Group classification of the class of variable-coefficient convection--diffusion equations was studied in~\cite{popo2004b,ivan2010a}:
\begin{gather*}
	f(x)u_t = (g(x)A(u)u_{x})_x + h(x)B(u)u_x.
\end{gather*}
An exhaustive analysis of Lie symmetries, up to equivalence transformations, was carried out for several subclasses of variable-coefficient nonlinear reaction--diffusion equations in~\cite{vaneeva-john-popo-soph,vaneeva-popo-soph-2009,vane2012b,vaneeva-zhalij}:
\begin{gather*}
f(x)u_t = (g(x)A(u)u_{x})_x + h(x)B(u).
\end{gather*}
In the last two equations, $u=u(t,x)\in\mathbb{R}^1$, $A=A(u)$, $B=B(u)$, $f=f(x)$, $g=g(x)$, and $h=h(x)$ are arbitrary smooth functions of their arguments satisfying $fgA(h^2+B^2)\ne0$.
The papers~\cite{abr-lah-sam,lah-sam,bashor-lah-zhdanov,zhdanov-lah,lah-spich-stoh} are devoted to the complete group classification of the most general class of quasilinear evolution equations
\begin{equation*}
	u_{t}=F(t,x,u,u_{x})u_{xx} + G(t,x,u,u_{x})
\end{equation*}
with arbitrary smooth functions $F=F(t,x,u,u_x)$ and $G=G(t,x,u,u_x)$, where $F\ne0$.
A complete group classification of reaction--convection--diffusion equations was obtained for one space variable in~\cite{cher-ser-rass,cher-ser-pliukh}, and for two space variables in~\cite{ser-serova-pryst,cher-ser-pryst}.
Symmetry properties of the class of reaction--diffusion systems
\begin{gather*}
	U_t=\varLambda U_{xx}+H(U)
\end{gather*}
with different constant nonzero diffusion matrices $\varLambda$ and an arbitrary nonzero smooth two-dimensional vector function $H=H(U)$ were investigated independently in~\cite{nikitin-2005,nikitin-2001} and~\cite{cher-king,fusch-cher}.
The Galilean invariance of the class of reaction--convection--diffusion systems
\begin{gather*}
		U_t=\p_x(F(U)U_x)+G(U)U_x+H(U),
\end{gather*}
where
\begin{gather*}
	U=\begin{pmatrix}
		u^1 \\
		u^2
	\end{pmatrix}
	\in \mathbb{R}^2, \quad F(U)=\begin{pmatrix}
		f^{11} & f^{12}\\
		f^{21} & f^{22}
	\end{pmatrix}, \quad  G(U)=\begin{pmatrix}
	g^{11} & g^{12}\\
	g^{21} & g^{22}
	\end{pmatrix}, \quad H(U)=\begin{pmatrix}
	h^1 \\
	h^2
	\end{pmatrix},
\end{gather*}
$f^{ab}=f^{ab}(U)$, $g^{ab}=g^{ab}(U)$, and $h^a=h^a(U)$ are arbitrary smooth functions, and $F(U)$ is a nonzero matrix; this invariance
was investigated in~\cite{ser-karp-pliukh-rass}; see also~\cite{ser-karp}.
The conformal invariance of class~\eqref{eq:mainsystem} was studied in~\cite{ser-ich,baik-gladk-wiltsh}.
Lie symmetries and exact solutions of diffusion-type equations and systems were studied in the PhD thesis of E.~Demetriou~\cite{demetr}. Group classification of systems with diagonal diffusion matrices was also studied in the technical report by N.~M.~Ivanova and C.~Sophocleous~\cite{ivan-soph-2011} and in the paper by S.~Kontogiorgis and C.~Sophocleous~\cite{kont-soph}.
Lie symmetries of a class of nonlinear diffusion systems in three space variables were studied in~\cite{stepanova-2017,stepanova-2019}.
The linearization problem and equivalence transformations for a class of nonlinear diffusion systems were investigated in~\cite{soph-wiltsh}.

Although more than half a century has passed since the publication of L.~V.~Ovsyannikov's paper~\cite{ovs-gf}, the problem of classifying the symmetry properties of nonlinear diffusion systems~\eqref{eq:mainsystem} remains unsolved in full generality. The difficulty is that, unlike scalar diffusion equations~\eqref{eq:nonlinear-ovs}, which contain only one arbitrary element $f$ depending on a single variable $u$, class~\eqref{eq:mainsystem} contains four arbitrary elements depending on two variables, $f^{ab}=f^{ab}(u^1,u^2)$.

\section{System of determining equations.\\ Kernel of maximal Lie invariance algebras}
	
Applying Lie's algorithm to system~\eqref{eq:mainsystem} (see, e.g.,~\cite{lie-1893,ovs-ga,olver-eng,lie-1881e,lie-1881d}), we obtain a system of equations determining the nonlinearities $f^{ab}=f^{ab}(u^1,u^2)$ and the coefficients of a Lie symmetry vector field of this system,
\begin{gather}\label{eq:inf-operator}
		X=\tau(t,x,u^1,u^2)\p_t+\xi(t,x,u^1,u^2)\p_x+\eta^a(t,x,u^1,u^2)\p_{u^a}.
\end{gather}
This system has the form (see also~\cite{ser-rass})
\begin{subequations}\label{eq:def-system}
	\begin{align}
	&\tau_x=\tau_{u^a}=\xi_{u^a}=0, \label{eq:def-system(a)}\\
	&\eta^c f^{ab}_{u^c}+\left(\tau_t-2\xi_{x}\right)f^{ab}+\eta^c_{u^b} f^{ac}-\eta^a_{u^c} f^{cb}=0, \label{eq:def-system(b)} \\
	&\eta^c_x\left(f^{ab}_{u^c}+f^{ac}_{u^b}\right)+2\eta^c_{xu^b}f^{ac}-\xi_{xx}f^{ab}+\delta_{ab}\xi_{t}=0,\\
	&\eta^a_t-\eta^b_{xx}f^{ab}=0,\\
	&f^{db}\eta^a_{u^cu^d}+f^{dc}\eta^a_{u^bu^d}=0, \label{eq:def-system(d)}
	\end{align}
\end{subequations}
where $\delta_{ab}$ is the Kronecker symbol.

In the determining system~\eqref{eq:def-system}, a superscript denotes the component number and a subscript denotes differentiation with respect to the corresponding variable.

We write system~\eqref{eq:def-system(d)} in expanded form:

\begin{gather}\label{eq:def-system(d)-matrix}
	\begin{pmatrix}
			f^{11} & f^{21} & 0\\
			f^{12} & f^{11}+f^{22} & f^{21}\\
			0 & f^{12} & f^{22}
		\end{pmatrix} \begin{pmatrix}
		\eta^a_{u^1u^1} \\
		\eta^a_{u^1u^2} \\
		\eta^a_{u^2u^2}
		\end{pmatrix}=0.
\end{gather}

Since~\eqref{eq:def-system(d)-matrix} is a system of homogeneous linear algebraic equations for the unknowns $\eta^a_{u^1u^1}$, $\eta^a_{u^1u^2}$, and $\eta^a_{u^2u^2}$, its solution depends on the determinant

	\begin{gather}\label{eq:def-system(d)-determinant}
\Delta=\left|\begin{matrix}
			f^{11} & f^{21} & 0\\
		f^{12} & f^{11}+f^{22} & f^{21}\\
		0 & f^{12} & f^{22}
	\end{matrix}\right|.
\end{gather}
Calculating determinant~\eqref{eq:def-system(d)-determinant}, we obtain

\begin{gather*}
	\Delta=(f^{11}+f^{22})(f^{11}f^{22}-f^{12}f^{21})=(f^{11}+f^{22})\left|\begin{matrix}
		f^{11} & f^{12}\\
		f^{21} & f^{22}
	\end{matrix}\right|=\tr F\cdot \det F.
\end{gather*}
If
\begin{gather}\label{eq:spur}
\Delta=\tr F\cdot \det F\ne0,
\end{gather}
then system~\eqref{eq:def-system(d)} has only the zero solution
	\begin{gather}\label{eq:solution}
\eta^a_{u^bu^c}=0.
\end{gather}
If $\Delta=0$, then system~\eqref{eq:def-system(d)} admits nonzero solutions in addition to~\eqref{eq:solution}.

Consider three simplest systems of class~\eqref{eq:mainsystem}:

\noindent 1) the system of heat conduction or diffusion equations
\[
\begin{array}{c}
	\begin{aligned}
	&u^1_t = u^1_{xx}, \\
	&u^2_t = u^2_{xx}
	\end{aligned}
\end{array}, \quad
F = \begin{pmatrix} 1 & 0 \\ 0 & 1 \end{pmatrix}, \quad \tr F \neq 0, \quad \det F \neq 0;
\]

\noindent 2) the system of equations for a two-phase liquid
\[
\begin{array}{c}
	\begin{aligned}
		&u^1_t=u^1_{xx},\\
		&u^2_t=u^1_{xx}
	\end{aligned}
\end{array},	
		\quad F=\begin{pmatrix}
			1 & 0\\
			1 & 0
		\end{pmatrix}, \quad \tr F\ne0, \quad \det F=0;
\]

\noindent 3) the Schr\"odinger equation ${\rm i}\psi_t=\psi_{xx}$ for the complex function $\psi=u^1+{\rm i}u^2$, or equivalently,
\[
\begin{array}{c}
	\begin{aligned}
	&u^1_t=u^2_{xx},\\
	&u^2_t=-u^1_{xx}
\end{aligned}
\end{array},	
\quad F=\begin{pmatrix}
	0 & 1\\
	-1 & 0
\end{pmatrix}, \quad \tr F=0, \quad \det F\ne0.
\]

In view of the above, we introduce the following definition.

 \begin{definition}
 We call a system of equations~\eqref{eq:mainsystem}

 1) a system of nonlinear diffusion equations if $\tr F\cdot\det F\ne0$;

 2) a system of nonlinear equations for a two-phase liquid if $\tr F\ne0$, $\det F=0$;

 3) a system of nonlinear Schr\"odinger-type equations if $\tr F=0$, $\det F\ne0$.
 \end{definition}	

We determine which systems of class~\eqref{eq:mainsystem} satisfying~\eqref{eq:spur} admit extensions of the kernel and carry out the full group classification of the first subclass.

From~\eqref{eq:def-system(a)} and~\eqref{eq:solution} we obtain
\begin{gather}\label{eq:coordinates-of-inf-operator}
		\tau=\tau(t), \quad	\xi=\xi(t,x), \quad \eta^a=\alpha^{ab}(t,x)u^b+\beta^a(t,x),
\end{gather}
where $\tau(t)$, $\xi(t,x)$, $\alpha^{ab}(t,x)$, and $\beta^a(t,x)$ are arbitrary smooth functions of their arguments.

We first determine the symmetries admitted by class~\eqref{eq:mainsystem} for arbitrary nonlinearities $f^{ab}=f^{ab}(u^1,u^2)$. Splitting the determining system~\eqref{eq:def-system} with respect to $f^{ab}$ and $f^{ab}_{u^c}$, we obtain
\begin{gather}\label{eq:sol}
		\xi_t=0, \quad \tau_t-2\xi_x=0, \quad \xi_{xx}=0, \quad \eta^a=0.
\end{gather}
The general solution of system~\eqref{eq:sol} is given by
\begin{gather}\label{eq:general-sol}
			\tau=2c_1t+c_2, \quad \xi=c_1x+c_3, \quad \eta^a=0,
\end{gather}
where $c_1$, $c_2$, and $c_3$ are constants of integration.

The vector field~\eqref{eq:inf-operator} with coefficients~\eqref{eq:general-sol} generates the Lie algebra
\begin{gather}\label{eq:kernel-of-mainsystem}
A^\cap=\langle\p_t, \, \p_x, \, D\rangle, \quad\text{where}\quad D:=2t\p_t+x\p_x,
\end{gather}
which is the kernel of maximal Lie invariance algebras of systems from class~\eqref{eq:mainsystem}.
For an individual system, we denote its maximal Lie invariance algebra by $A^{\max}$.

\section{Equivalence transformations}

Knowledge of equivalence transformations greatly facilitates the group classification of a class of differential equations. These transformations allow us to divide a class into inequivalent subclasses, identifying canonical systems in each subclass. It is sufficient to study canonical representatives of each subclass to determine the symmetry properties of all systems from class~\eqref{eq:mainsystem}.

We seek the equivalence transformations of class~\eqref{eq:mainsystem}
and investigate, up to these transformations, the symmetry properties of its systems.

Continuous equivalence transformations of class~\eqref{eq:mainsystem} are found by the algorithm
proposed in~\cite{ahat-gaz-ibr}.
The following theorem holds.

\begin{theorem}\label{thm:equivalence-algebra-of-the-main-system}
The equivalence algebra of class~\eqref{eq:mainsystem} is the linear span of the vector fields

\[
\p_t,\ \p_x,\ \p_{u^a},\ t\p_t-f^{ab}\p_{f^{ab}},\ x\p_x+2f^{ab}\p_{f^{ab}},\
u^a\p_{u^b}+f^{ac}\p_{f^{bc}}-f^{cb}\p_{f^{ca}}
\]
in the space $\mathbb R^2_{t,x}\times\mathbb R^2_U\times\mathbb R^4_F$.
This equivalence algebra is isomorphic to
$\mathfrak{aff}(1,\mathbb R)\oplus\mathfrak{aff}(1,\mathbb{R})\oplus\mathfrak{aff}(2,\mathbb{R})$.
\end{theorem}

\begin{proof}
 Applying the algorithm for finding equivalence transformations proposed in~\cite{ahat-gaz-ibr}, we obtain the determining equations for the coefficients of the equivalence vector field
	\begin{gather}\label{eq:operator-of-equivalence}
		E=\tau(t,x,U)\p_t+\xi(t,x,U)\p_x+\eta^a(t,x,U)\p_{u^a}+\zeta^{ab}(t,x,U,F)\p_{f^{ab}}.
	\end{gather}
 This system has the form
	\begin{gather}\label{eq:system-of-equivalence}
		\begin{aligned}
			&\tau_{tt}=\tau_x=\tau_{u^a}=\xi_t=\xi_{xx}=\xi_{u^a}=\eta^a_t=\eta^a_x=\eta^a_{u^bu^c}=0,\\
			&\zeta^{ab}=(2\xi_x-\tau_t)f^{ab}+\eta^a_{u^c}f^{cb}-\eta^c_{u^b}f^{ac}.
		\end{aligned}
	\end{gather}
 The general solution of system~\eqref{eq:system-of-equivalence} is
	\begin{gather}\label{eq:gensol-of-equivalence}
			\tau=c_1t+d_0,\quad
			\xi=c_2x+d_1,\quad
			\eta^a=k_{ab}u^b+m_a,\quad
			\zeta^{ab}=(2c_2-c_1)f^{ab}+k_{ac}f^{cb}-k_{cb}f^{ac},
	\end{gather}
 where $c_1$, $c_2$, $d_0$, $d_1$, $k_{ab}$, and $m_a$ are arbitrary constants.
Substituting~\eqref{eq:gensol-of-equivalence} into~\eqref{eq:operator-of-equivalence}, we collect the coefficients of the arbitrary constants.
We thereby obtain the basis vector fields stated in the theorem.
\end{proof}

\begin{corollary}
The group of continuous equivalence transformations of class~\eqref{eq:mainsystem} consists of the point transformations

\begin{gather}\label{eq:inf-transform-of-equivalence}
\tilde t=l_0t+s_0, \quad \tilde x=l_1x+s_1, \quad \tilde U=PU+Q, \quad \tilde F=\frac{l^2_1}{l_0}PFP^{-1}
\end{gather}
in the space $\mathbb R^2_{t,x}\times\mathbb R^2_U\times\mathbb R^4_F$,
where
\begin{gather*}
	   P=\begin{pmatrix}
			p_{11} & p_{12}\\
			p_{21} & p_{22}
		 \end{pmatrix}, \quad
		Q=\begin{pmatrix}
			q_1 \\
			q_2
		 \end{pmatrix},
\end{gather*}
$l_0$, $l_1$, $s_0$, $s_1$, $p_{ab}$, and $q_a$ are arbitrary constants satisfying $l_0>0$, $l_1>0$, and $\det P>0$.
This group is isomorphic to
$\mathrm{Aff}^+(1,\mathbb R)\times\mathrm{Aff}^+(1,\mathbb R)\times\mathrm{Aff}^+(2,\mathbb R)$.
\end{corollary}

\begin{remark}\label{rem:EquivGroup}
In addition to continuous transformations~\eqref{eq:inf-transform-of-equivalence}, class~\eqref{eq:mainsystem}
admits the following discrete equivalence transformations:
\begin{gather*}
\begin{aligned}
&\tilde t=-t,\ \tilde x=x, \ \tilde u^a=u^a,\ \tilde f^{ab}=-f^{ab},\\
&\tilde t=t ,\ \tilde x=-x,\ \tilde u^a=u^a,\ \tilde f^{ab}=f^{ab},\\
&\tilde t=t ,\ \tilde x=x, \ \tilde u^1=-u^1,\ \tilde u^2 =u^2,\
 \tilde f^{11}=f^{11},\ \tilde f^{12}=-f^{12},\ \tilde f^{21}=-f^{21},\ \tilde f^{22}=f^{22}.\\
\end{aligned}
\end{gather*}
For completeness, consider a general point equivalence transformation. Splitting the transformed system with respect to the highest-order derivatives first gives $\tilde t=T(t)$, $\tilde x=X(t,x)$ and $\tilde U=\Psi(t,x,U)$. The coefficients of $U_{xx}$ give $\tilde F=(X_x^2/T_t)\Psi_U F\Psi_U^{-1}$. Since $F(U)$ is arbitrary, splitting the terms quadratic in $U_x$ yields $\Psi_{UU}=0$, while the remaining terms imply $T_{tt}=X_t=X_{xx}=0$ and $\Psi_t=\Psi_x=0$. Thus every equivalence transformation has the affine form~\eqref{eq:inf-transform-of-equivalence}. The three reflections above account for the independent signs of $l_0$, $l_1$, and $\det P$.
These transformations exhaust the complete list of discrete equivalence transformations of class~\eqref{eq:mainsystem}
that are independent, up to compositions with each other and with continuous equivalence transformations.
The complete equivalence group $G^\sim$ of class~\eqref{eq:mainsystem}
consists of transformations~\eqref{eq:inf-transform-of-equivalence}, where
$l_0$, $l_1$, $s_0$, $s_1$, $p_{ab}$, and $q_a$ are arbitrary constants such
that $l_0l_1\ne0$, and $P$ is an arbitrary constant nonsingular matrix.
This group is isomorphic to
$\mathrm{Aff}(1,\mathbb R)\times\mathrm{Aff}(1,\mathbb R)\times\mathrm{Aff}(2,\mathbb R)$.
The affine equivalence transformations~\eqref{eq:inf-transform-of-equivalence} were also given in a different notation in~\cite[Sec.~2.1]{soph-wiltsh}.
\end{remark}

\section{Necessary and sufficient conditions for extensions\\ of the kernel of maximal Lie invariance algebras}

We derive necessary and sufficient conditions for an extension of the kernel of maximal Lie invariance algebras of the class of nonlinear diffusion systems~\eqref{eq:mainsystem}. In other words, we determine the nonlinearities $f^{ab}$ for which a system from this class is invariant under a Lie algebra larger than~\eqref{eq:kernel-of-mainsystem}.

\begin{theorem}\label{thm:necessary-and-sufficient}
A system from class~\eqref{eq:mainsystem} admits an extension of the kernel of maximal Lie invariance algebras~\eqref{eq:kernel-of-mainsystem} if and only if, up to $G^\sim$-equivalence,
the nonlinearities $f^{ab}=f^{ab}(u^1,u^2)$ have one of the forms listed in Table~\ref{tab1}.
\end{theorem}

\begingroup\small
\begin{longtable}{|c|l|l|}
\captionsetup{width=\textwidth}
\caption{Forms of diffusion coefficients for which systems of class~\eqref{eq:mainsystem} admit extensions of the kernel of maximal Lie invariance algebras} \label{tab1} \\
	\hline
 No. &$f^{ab}$ & Kernel, conditions \\
	\hline
	\endfirsthead
	\hline
 No. &$f^{ab}$ & Kernel, conditions \\
	\hline
	\endhead
	\hline
	\endfoot
	1    & $f^{ab}={\rm e}^{nu^1}\varphi^{ab}$, \ $\omega=u^2$    & $	A^\cap_1=A^\cap\rsemioplus\langle nt\p_t-\p_{u^1} \rangle$, \ $n\in\{0,1\}$\\ \hline
	2    & $f^{11}={\rm e}^{n\frac{u^1}{u^2}}\bigl(\varphi^{11}+\frac{u^1}{u^2}\varphi^{21}\bigr)$,    & $	A^\cap_2= A^\cap\rsemioplus \langle nt\p_t-u^2\p_{u^1}\rangle$, \ $n\in \{0,1\}$  \\
	& $f^{12}={\rm e}^{n\frac{u^1}{u^2}}\bigl (\varphi^{12}+\frac{u^1}{u^2}\left(\varphi^{22}-\varphi^{11}\right)-\bigl(\frac{u^1}{u^2}\bigr)^2\varphi^{21}\bigr)$, &
	\\
	& $f^{21}={\rm e}^{n\frac{u^1}{u^2}}\varphi^{21}$, \ $f^{22}={\rm e}^{n\frac{u^1}{u^2}}\bigl(\varphi^{22}-\frac{u^1}{u^2}\varphi^{21}\bigr)$, \ $\omega=u^2$&  \\
	\hline
	3 & $f^{11}=(u^1)^n\varphi^{11}$, \ $f^{12}=(u^1)^{n+1}\varphi^{12}$, & $A^\cap_3= A^\cap\rsemioplus \langle nt\p_t-u^1\p_{u^1}\rangle$, \ $n\in \mathbb{R}$ \\
	& $f^{21}=(u^1)^{n-1}\varphi^{21}$, \ $f^{22}=(u^1)^n\varphi^{22}$, \ $\omega=u^2$  & \\
	\hline
	4 & $f^{11}=\left(u^2\right)^n\varphi^{11}$, \ $f^{12}=\left(u^2\right)^{n-1}\varphi^{12}$, &$A^\cap_4= A^\cap\rsemioplus\big\langle nt\p_t-u^2\p_{u^2}-\p_{u^1}\big\rangle$, \ $n\in \mathbb{R}$ \\
	& $f^{21}=\left(u^2\right)^{n+1}\varphi^{21}$, \ $f^{22}=\left(u^2\right)^n\varphi^{22}$, \ $\omega=u^1-\ln u^2$   &  \\
	\hline
	5 & $f^{11}={\rm e}^{nu^1}(\varphi^{11}-u^1\varphi^{12})$,\quad $f^{12}={\rm e}^{nu^1}\varphi^{12}$,
      &$A^\cap_5= A^\cap\rsemioplus\big\langle nt\p_t-u^1\p_{u^2}-\p_{u^1}\big\rangle$,  \\
	& $f^{21}={\rm e}^{nu^1}\big(\varphi^{21}+u^1(\varphi^{11}-\varphi^{22})-(u^1)^2\varphi^{12}\big)$,
    & $n\in\{0,1\}$  \\
	& $f^{22}={\rm e}^{nu^1}(\varphi^{22}+u^1\varphi^{12})$, \ $\omega=u^2-\frac12(u^1)^2$ &  \\
	\hline
	6 & $f^{11}=(u^1)^n\varphi^{11}$, \ $f^{12}=(u^1)^{n-m}\varphi^{12}$, & $A^\cap_6= A^\cap\rsemioplus\langle nt\p_t-I-mu^2\p_{u^2}\rangle$, \\
	& $f^{21}=(u^1)^{n+m}\varphi^{21}$, \ $f^{22}=(u^1)^n\varphi^{22}$, \ $\omega=(u^1)^{-1-m}u^2$ &$n\in \mathbb{R}$, \, $m\in\mathbb{R}$, \, $m\ne-1,0$  \\
	\hline
	7 & $f^{ab}=(u^1)^n\varphi^{ab}$, \ $\omega=u^2/u^1$ &
$A^\cap_7= A^\cap\rsemioplus\langle nt\p_t-I\rangle$, \ $n\in \mathbb{R}$ \\
	\hline
	8 & $f^{11}=\left(u^2\right)^n\left(\varphi^{11}+\varphi^{21}\ln u^2\right)$, & $A^\cap_8= A^\cap\rsemioplus\langle nt\p_t-I-u^2\p_{u^1}\rangle$, \\
	& $f^{12}=\left(u^2\right)^n\left(\varphi^{12}+\left(\varphi^{22}-\varphi^{11}\right)\ln u^2-\varphi^{21}\ln^2 u^2\right)$, &$n\in \mathbb{R}$  \\
	& $f^{21}=\left(u^2\right)^n\varphi^{21}$, &  \\
	& $f^{22}=\left(u^2\right)^n\left(\varphi^{22}-\varphi^{21}\ln u^2\right)$, \ $\omega=u^1/u^2-\ln u^2$  &  \\
	\hline
	9 & $f^{1a}={\rm e}^{-n\theta}\bigl(2\rho^{-2}u^a \vec u \cdot \vec\varphi+\psi^a\bigr)$, & $A^\cap_9= A^\cap\rsemioplus\langle nt\p_t-mI-J\rangle$, \\
	& $f^{2a}={\rm e}^{-n\theta}\bigl(2\rho^{-2}u^a\vec u \cdot \vec\varphi^\perp+(\vec\psi^\perp)^a
	\bigr)$, \ $\omega=\rho{\rm e}^{m\theta}$
 &$n\in \mathbb{R}$, \, $m\in \mathbb{R}$ \\
	\hline
\noprint{	
9 & $f^{1a}={\rm e}^{n\theta}\bigl(\rho^{-2}u^a u^b\varphi^b+\psi^a\bigr)$, & $A^\cap_9=\langle \p_t, \, \p_x, \, D$, &$n\in \mathbb{R}$, \\
	& $f^{2a}={\rm e}^{n\theta}\bigl(\rho^{-2}u^a\varepsilon_{bc}u^c\varphi^b-\varepsilon_{ab}\psi^b\bigr)$, \
$\omega=\rho{\rm e}^{m\arctan(u^2/u^1)}$ &$nt\p_t-(mI+J)\rangle$ &$m\in \mathbb{R}$ \\
\hline
}
\end{longtable}
\endgroup

Throughout cases 1--8 of Table~\ref{tab1}, $\varphi^{ab}=\varphi^{ab}(\omega)$ are arbitrary smooth functions satisfying $\varphi^{11}+\varphi^{22}\ne0$ and $\varphi^{11}\varphi^{22}-\varphi^{12}\varphi^{21}\ne0$, and $I=u^1\p_{u^1}+u^2\p_{u^2}$, $J=u^2\p_{u^1}-u^1\p_{u^2}$.
In case 9 of Table~\ref{tab1},
$\vec u=(u^1,u^2)$,
$\vec\varphi=(\varphi^1,\varphi^2)$,
$\vec\psi=(\psi^1,\psi^2)$,
$\vec\varphi^\perp=(-\varphi^2,\varphi^1)$,
$\vec\psi^\perp=(-\psi^2,\psi^1)$,
$\varphi^a=\varphi^a(\omega)$,
$\psi^a=\psi^a(\omega)$ are arbitrary smooth functions,
$\varphi^1\ne-\psi^1$,
$\lVert\vec\psi\rVert^2+2\vec\varphi\cdot\vec\psi\ne0$,
$\rho=\sqrt{(u^1)^2+(u^2)^2}$,
$\theta=\arctan(u^2/u^1)$ on a local chart with $u^1>0$.

We call the subclass of class~\eqref{eq:mainsystem} with diffusion coefficients listed in case~$N$ of Table~\ref{tab1}
the $N$th class of diffusion systems, $N\in\{1,\dots,9\}$.
It is then natural to call the subclass of class~\eqref{eq:mainsystem} whose systems admit no extension of the kernel of Lie symmetries $A^{\cap}$ the zero class.

\begin{proof}
We analyze subsystem~\eqref{eq:def-system(b)} of determining system~\eqref{eq:def-system}, rewriting it with the aid of~\eqref{eq:coordinates-of-inf-operator} as follows:
	\begin{gather}\label{eq:second-of-def-system}
		\left(\alpha^{cd}u^d+\beta^c\right) f^{ab}_{u^c}+\left(\tau_t-2\xi_{x}\right)f^{ab}+\alpha^{cb} f^{ac}-\alpha^{ac} f^{cb}=0.
	\end{gather}
If the coefficient functions $\alpha^{cd}$ and $\beta^c$ vanish identically for every
Lie symmetry vector field $X$ of a given system from class~\eqref{eq:mainsystem},
then this system admits no extension of the kernel of Lie symmetries~\eqref{eq:kernel-of-mainsystem}.
Consequently, an extension requires that at least one of the coefficient functions $\alpha^{cd}$ and $\beta^c$
be nonzero for some value of $(t,x)$.
Fixing such a pair of values in~\eqref{eq:second-of-def-system}, we obtain
	\begin{gather}\label{eq:structure-system}
		(k_{cd}u^d+m_c) f^{ab}_{u^c}-nf^{ab}+k_{cb} f^{ac}-k_{ac} f^{cb}=0,
	\end{gather}
where $k_{ab}$, $m_a$, and $n$ are constants, which will be called \textit{structural constants}.
We call~\eqref{eq:structure-system} a \textit{structural system} for determining the functions $f^{ab}$.\footnote{Fixing $(t,x)$ in the determining equations is a step in the method of \emph{furcate splitting}. The resulting equations for arbitrary elements are called \emph{template-form equations}; see~\cite{opan2020b}. The furcate splitting method was first applied in~\cite{niki2001a}, named in~\cite{ivan2010a}, and formalized in~\cite{opan2020b}; for applications, see~\cite{bihl2020a,popo2004b,vane2012b,vane2015a,vaneeva-zhalij}.}

Conversely, if $Y=(KU+M)\cdot\partial_U$ satisfies
$Y(F)=nF+[K,F]$ with constant $K,M,n$, then
$Y-nt\p_t$ satisfies the full determining system:
the $t,x$-derivatives of $K,M$ vanish because these coefficients are constant,
while the second derivatives of $Y$ with respect to $U$ vanish because $Y$ is affine.
Thus every solution of the structural system indeed yields a Lie symmetry.

By construction, not all of the constants $k_{ab}$ and $m_a$ vanish.
	
We solve structural system~\eqref{eq:structure-system} up to $G^\sim$-equivalence. The characteristic system associated with equations~\eqref{eq:structure-system} is
	\begin{gather}\label{eq:characteristic-system}
		\dfrac{\mathrm{ d}u^a}{k_{ab}u^b+m_a}=\dfrac{\mathrm{ d}f^{ab}}{nf^{ab}-k_{cb}f^{ac}+k_{ac}f^{cb}}=\mathrm{ d}p,
	\end{gather}
where $p$ is an auxiliary parameter to be eliminated after integration.
	
The characteristic system~\eqref{eq:characteristic-system} splits into two independent subsystems
\begin{gather}\label{eq:first-system-of-characteristic-system}
		\dot u^a=k_{ab}u^b+m_a,
       \\\label{eq:second-system-of-characteristic-system}
		\dot f^{ab}=nf^{ab}-k_{cb}f^{ac}+k_{ac}f^{cb}.
	\end{gather}
A dot above a function denotes differentiation with respect to $p$, that is,
$\dot u^a:=\mathrm du^a/\mathrm dp$,
$\dot f^{ab}:=\mathrm df^{ab}/\mathrm dp$.
System~\eqref{eq:first-system-of-characteristic-system} can be written in matrix form,
	\begin{gather*}
 \dot U=KU+M,\quad\text{where}\quad
K:=\begin{pmatrix}
		k_{11} & k_{12}\\
		k_{21} & k_{22}
	\end{pmatrix}, \ \
M:=\begin{pmatrix}
		m_1 \\
		m_2
	\end{pmatrix}	
\ \ \text{are constant matrices.}
    \end{gather*}

The action of the equivalence group $G^\sim$ on class~\eqref{eq:mainsystem} induces an equivalence relation
on systems of the form~\eqref{eq:first-system-of-characteristic-system};
this coincides with the equivalence relation
induced by the action of the affine group in the space with coordinates $(u^1,u^2)$
on these systems.
This group consists of affine transformations $\tilde U=PU+Q$,
where $P$ is an arbitrary constant nonsingular $2\times2$ matrix
and $Q$ is an arbitrary constant two-component column vector.
\par\noindent\begin{minipage}{\linewidth}
The canonical forms of~\eqref{eq:first-system-of-characteristic-system} are classified according to $\operatorname{rank}K$. This yields nine $G^\sim$-inequivalent subclasses of class~\eqref{eq:mainsystem}.
Here, the nine subclasses are distinguished by the nine inequivalent canonical types of additional Lie symmetry generators. The corresponding families of systems need not be disjoint, and individual systems from different families may be $G^\sim$-equivalent.
\end{minipage}

We also identify nonzero constant multiples of an affine generator,
since they span the same one-dimensional Lie algebra; this accounts
for the normalization of nonzero eigenvalues below.

In cases~1, 2, and~5 of Table~\ref{tab1}, the structural constant $n$ can also be normalized.
Indeed, multiplying $(K,M,n)$ by a common nonzero constant does not change
structural system~\eqref{eq:structure-system}. If $n\ne0$, multiply these
constants by $1/n$ and apply the linear changes
$(\tilde u^1,\tilde u^2)=(nu^1,u^2)$ in cases~1 and~2, and
$(\tilde u^1,\tilde u^2)=(nu^1,n^2u^2)$ in case~5.
These changes restore the corresponding canonical forms of $K,M$ and
reduce $n$ to~$1$. Hence $n\in\{0,1\}$ in these three cases.
For the other canonical types, this rescaling would alter the chosen
normalization of $K$, so $n$ remains an arbitrary real constant.

\medskip\par\noindent
\textbf{1}. $\mathop{\rm rank}K=0$. This case occurs only for $K=0$;
then $M\ne0$ and~\eqref{eq:first-system-of-characteristic-system} takes the form
$\dot U=M$. 
Linear transformations of $U$ reduce the column vector $M$ to $M=(1,0)^{\mathsf T}$.
Indeed, up to interchanging $u^1$ and $u^2$, we may assume that $m_1\ne0$.
Then it suffices to apply $u^1\rightarrow u^1/m_1$, $u^2\rightarrow u^2-m_2u^1/m_1$.

\medskip\par\noindent
\textbf{2}. $\mathop{\rm rank}K=1$. Without loss of generality, we assume that the rows of $K$ are proportional
and system~\eqref{eq:first-system-of-characteristic-system} has the form
	\begin{gather}\label{eq:system-det0}
		\dot u^1=k\left(k_{21}u^1+k_{22}u^2\right)+m_1, \quad
		\dot u^2=k_{21}u^1+k_{22}u^2+m_2,
	\end{gather}
where $k$ is a constant. Inequivalent subclasses of systems~\eqref{eq:system-det0} are distinguished by the values of $\tilde m_1:=m_1-km_2$ and $\tilde k_{22}:=k_{22}+kk_{21}$.
The corresponding canonical forms and affine transformations are as follows.

\medskip\par\noindent
2.1. \ $\tilde m_1=0$, \ $\tilde k_{22}=0$, \ $k_{21}\ne0$ \ $(k_{21}=1)$; \
$u^1\rightarrow u^1-ku^2+m_2$, \ $u^2\rightarrow u^2$; \
	\begin{gather}\label{eq:IIclass}
\dot U=\begin{pmatrix}
		0 & 0\\
		1 & 0
	\end{pmatrix}U.
\end{gather}

\noindent
2.2. \ $\tilde m_1=0$, \ $\tilde k_{22}\ne0$ \ $(\tilde k_{22}=1)$; \
$u^1\rightarrow u^1-ku^2$, \
$u^2\rightarrow k_{21}u^1+k_{22}u^2+m_2$;
		\begin{gather}\label{eq:IIIclass}
	\dot U=\begin{pmatrix}
			0 & 0\\
			0 & 1
		\end{pmatrix}U.
	\end{gather}

\noindent
2.3. \ $\tilde m_1\ne0$, \ $\tilde k_{22}=0$, \ $k_{21}\ne0$, \ $(k_{21}=1)$; \
$u^1\rightarrow (u^1-ku^2+m_2)/\tilde m_1$, \
$u^2\rightarrow u^2/\tilde m_1$;
		\begin{gather}\label{eq:IVclass}
	\dot U=\begin{pmatrix}
			0 & 0\\
			1 & 0
		\end{pmatrix}U+\begin{pmatrix}
			1 \\
			0
		\end{pmatrix}.
	\end{gather}

\noindent
	2.4. $\tilde m_1\ne0$, \ $\tilde k_{22}\ne0$, \  $(\tilde k_{22}=1)$; \
$u^1\rightarrow (u^1-ku^2)/\tilde m_1$, \
$u^2\rightarrow k_{21}(u^1+m_1)+k_{22}(u^2+m_2)$;
	\begin{gather}\label{eq:Vclass}
	\dot U=\begin{pmatrix}
			0 & 0\\
			0 & 1
		\end{pmatrix}U+\begin{pmatrix}
			1 \\
			0
		\end{pmatrix}.
	\end{gather}

\par\noindent
\textbf{3}. $\mathop{\rm rank}K=2$.
The affine equivalence transformation $u^a\rightarrow u^a+q_a$ reduces
system~\eqref{eq:first-system-of-characteristic-system} to
	\begin{gather}\label{eq:system-detne0}
		\dot u^a=k_{ab}u^b-k_{ab}q_b+m_a.
	\end{gather}
Since $\det K\ne0$, there exist parameters $q_a$ such that $k_{ab}q_b=m_a$,
so system~\eqref{eq:system-detne0} takes the form of a system of homogeneous linear ordinary differential equations
with constant coefficients:
	\begin{gather}\label{eq:system-detne0-linear}
		\dot U=KU.
	\end{gather}
	
By means of linear transformations of $U$, system~\eqref{eq:system-detne0-linear} can be reduced to a form in which $K$ is one of the following real Jordan canonical forms:
\begin{gather*}
		K^{\mathrm{I}}=\begin{pmatrix}
			1 & 0\\
			0 & 1+m
		\end{pmatrix}, \quad
		K^{\mathrm{II}}=\begin{pmatrix}
			1 & 0\\
			0 & 1
		\end{pmatrix}, \quad
		K^{\mathrm{III}}=\left(\begin{matrix}
			1 & 1\\
			0 & 1
		\end{matrix}\right), \quad
		K^{\mathrm{IV}}=\left(\begin{matrix}
			s & 1\\
			-1 & s
		\end{matrix}\right),
	\end{gather*}
where $m$ and $s$ are arbitrary constants with $m\ne-1,0$.
	
As a result, we obtain the following four canonical types of systems of the form~\eqref{eq:first-system-of-characteristic-system}, up to affine transformations of $U$ and nonzero rescalings of the characteristic parameter:
	\begin{gather*}
	\dot U=K^{\mathrm{I}}U, \quad
	\dot U=K^{\mathrm{II}}U, \quad
	\dot U=K^{\mathrm{III}}U, \quad
	\dot U=K^{\mathrm{IV}}U. \quad
	\end{gather*}
Different parameter values within the same type may still be equivalent.
We find solutions of systems~\eqref{eq:first-system-of-characteristic-system} and~\eqref{eq:second-system-of-characteristic-system} for each of the above canonical cases, enumerated consecutively below.
	
	1. $k_{ab}=0$, $m_1=1$, $m_2=0$.
Systems~\eqref{eq:first-system-of-characteristic-system} and~\eqref{eq:second-system-of-characteristic-system} have the form
$\dot u^1=1$, $\dot u^2=0$, and $\dot f^{ab}=nf^{ab}$, and their general solutions are, respectively,
\begin{gather}\label{eq:I-class-gensol}
\begin{split}
&u^1=p+c_1,\quad
u^2=c_2,\\
&f^{ab}=\lambda_{ab}{\rm e}^{np},
\end{split}
\end{gather}
where $c_1$, $c_2$, and $\lambda_{ab}$ are constants of integration. Eliminating $p$ from~\eqref{eq:I-class-gensol}, we obtain the first integrals of characteristic system~\eqref{eq:characteristic-system}:
	\begin{gather*}
		I_1=u^2, \quad I_{ab}={\rm e}^{-nu^1}f^{ab}.
	\end{gather*}
Up to $G^\sim$-equivalence, we may assume that $n\in\{0,1\}$.
Then the general solution of structural system~\eqref{eq:structure-system} is given by the functions listed in case 1 of Table~\ref{tab1}.
	
2. To obtain the same first integral $I_1$ as in the preceding case, we interchange $u^1$ and $u^2$ in system~\eqref{eq:IIclass}. This yields $k_{11}=k_{21}=k_{22}=0$, $k_{12}=1$, and $m_1=m_2=0$.
We rewrite systems~\eqref{eq:first-system-of-characteristic-system} and~\eqref{eq:second-system-of-characteristic-system} as
	\begin{gather}\label{eq:II-class-for-u}
		\begin{split}
		&\dot u^1=u^2,\quad
		\dot u^2=0,\\
		&\dot f^{11}=nf^{11}+f^{21},\quad
			\dot f^{12}=nf^{12}-f^{11}+f^{22},\quad
			\dot f^{21}=nf^{21}, \quad
			\dot f^{22}=nf^{22}-f^{21}.
			\end{split}
		\end{gather}
The general solution of~\eqref{eq:II-class-for-u} is
	\begin{gather*}
		\begin{split}
		&u^1=c_1+c_2p,\quad
		u^2=c_2,\\
			&f^{11}=\left(\lambda_{11}+\lambda_{21}p\right){\rm e}^{np},\quad
			f^{12}=\left[\lambda_{12}+\left(\lambda_{22}-\lambda_{11}\right)p-\lambda_{21}p^2\right]{\rm e}^{np},\\
			&f^{21}=\lambda_{21} {\rm e}^{np},\quad
			f^{22}=\left(\lambda_{22}-\lambda_{21}p\right){\rm e}^{np}.
		\end{split}
	\end{gather*}
Up to $G^\sim$-equivalence, we may assume that $n\in\{0,1\}$.
After eliminating $p$, we obtain the functions $f^{ab}$ given in case 2 of Table~\ref{tab1}.
	
3. As in the preceding case, interchanging $u^1$ and $u^2$ in system~\eqref{eq:IIIclass} gives $k_{11}=1$, $k_{12}=k_{21}=k_{22}=m_1=m_2=0$.
Systems~\eqref{eq:first-system-of-characteristic-system} and~\eqref{eq:second-system-of-characteristic-system} take the form
	\begin{gather}\label{eq:III-class-for-u}
		\begin{split}
		&\dot u^1=u^1,\quad
		\dot u^2=0,\\
		&\dot f^{11}=nf^{11}, \quad \dot f^{12}=(n+1)f^{12}, \quad
		\dot f^{21}=(n-1)f^{21}, \quad \dot f^{22}=nf^{22}.
		\end{split}
	\end{gather}
The general solution of~\eqref{eq:III-class-for-u} is given by
\begin{gather}\label{eq:III-class-gensol-inter}
	\begin{split}
		&u^1=c_1{\rm e}^p,\quad
		u^2=c_2,\\
		&f^{11}=\lambda_{11}{\rm e}^{np}, \quad f^{12}=\lambda_{12}{\rm e}^{(n+1)p}, \quad
		f^{21}=\lambda_{21}{\rm e}^{(n-1)p}, \quad f^{22}=\lambda_{22}{\rm e}^{np},
	\end{split}
	\end{gather}
where $c_1$, $c_2$, and $\lambda_{ab}$ are constants of integration. Eliminating $p$ from~\eqref{eq:III-class-gensol-inter}, we obtain the general solution of structural system~\eqref{eq:structure-system}, given in case 3 of Table~\ref{tab1}.
	
4. From system~\eqref{eq:Vclass}, we have $k_{11}=k_{12}=k_{21}=0$, $k_{22}=1$, and $m_1=1$, $m_2=0$.
We write systems~\eqref{eq:first-system-of-characteristic-system} and~\eqref{eq:second-system-of-characteristic-system} as
\begin{gather}\label{eq:IV-class-for-u}
	\begin{split}
		&\dot u^1=1,\quad
		\dot u^2=u^2,\\
		&\dot f^{11}=nf^{11}, \quad \dot f^{12}=(n-1)f^{12}, \quad
		\dot f^{21}=(n+1)f^{21}, \quad \dot f^{22}=nf^{22}.
	\end{split}
	\end{gather}
The general solution of~\eqref{eq:IV-class-for-u} is
	\begin{align*}
		&u^1=p+c_1,\quad
		u^2=c_2{\rm e}^p,
		\\
		&f^{11}=\lambda_{11}{\rm e}^{np}, \quad f^{12}=\lambda_{12}{\rm e}^{(n-1)p}, \quad
		f^{21}=\lambda_{21}{\rm e}^{(n+1)p}, \quad f^{22}=\lambda_{22}{\rm e}^{np},
	\end{align*}
where $c_1$, $c_2$, and $\lambda_{ab}$ are constants of integration. Eliminating $p$, we obtain the general solution of structural system~\eqref{eq:structure-system}, which agrees with case 4 of Table~\ref{tab1}.
	
5. From system~\eqref{eq:IVclass}, we have $k_{11}=k_{12}=k_{22}=0$, $k_{21}=1$, and $m_1=1$, $m_2=0$.
Systems~\eqref{eq:first-system-of-characteristic-system} and~\eqref{eq:second-system-of-characteristic-system} take the form
	\begin{gather}\label{eq:V-class-for-u}
	\begin{split}
		&\dot u^1=1,\quad
		\dot u^2=u^1,\\
		&\dot f^{11}=nf^{11}-f^{12}, \quad \dot f^{12}=nf^{12}, \quad
		\dot f^{21}=nf^{21}+f^{11}-f^{22}, \quad \dot f^{22}=nf^{22}+f^{12}.
	\end{split}
	\end{gather}
The general solution of~\eqref{eq:V-class-for-u} is
	\begin{gather}\label{eq:V-class-gensol-inter}
		\begin{aligned}
			&u^1=p+c_1,\quad
			u^2=\frac{1}{2}\left(p+c_1\right)^2+c_2,\\
			&f^{11}=\left(\lambda_{11}-\lambda_{12}p\right){\rm e}^{np}, \quad
			f^{12}=\lambda_{12}{\rm e}^{np},\\
			&f^{21}=\left(\lambda_{21}+(\lambda_{11}-\lambda_{22})p-\lambda_{12}p^2\right){\rm e}^{np},\quad
			f^{22}=\left(\lambda_{22}+\lambda_{12}p\right){\rm e}^{np},
		\end{aligned}
	\end{gather}
where $c_1$, $c_2$, and $\lambda_{ab}$ are constants of integration. Up to $G^\sim$-equivalence, we may assume that $n\in\{0,1\}$. Eliminating $p$ from~\eqref{eq:V-class-gensol-inter}, we obtain the general solution of structural system~\eqref{eq:structure-system}, given in case 5 of Table~\ref{tab1}.
	
6. $k_{11}=1$,  $k_{12}=k_{21}=0$,  $k_{22}=1+m$,  $m_a=0$, $m \ne -1,0$.
Systems~\eqref{eq:first-system-of-characteristic-system} and~\eqref{eq:second-system-of-characteristic-system} are written as
	\begin{gather}\label{eq:VI-class-for-uf}
		\begin{aligned}
			\dot u^1&=u^1, \quad
			\dot u^2=(1+m)u^2,\\
			\dot f^{11}&=nf^{11}, \quad
			\dot f^{12}=(n-m)f^{12},\quad
			\dot f^{21}=(n+m)f^{21}, \quad
			\dot f^{22}=nf^{22}.
		\end{aligned}
	\end{gather}
The general solution of~\eqref{eq:VI-class-for-uf} is
	\begin{gather}\label{eq:VI-class-gensol-inter}
		\begin{aligned}
			u^1&=c_1{\rm e}^p,\quad
			u^2=c_2{\rm e}^{(1+m)p},\\
			f^{11}&=\lambda_{11}{\rm e}^{np},\quad
			f^{12}=\lambda_{12}{\rm e}^{(n-m)p},\quad
			f^{21}=\lambda_{21}{\rm e}^{(n+m)p},\quad
			f^{22}=\lambda_{22}{\rm e}^{np},
		\end{aligned}
	\end{gather}
where $c_1$, $c_2$, and $\lambda_{ab}$ are constants of integration. Eliminating $p$ from~\eqref{eq:VI-class-gensol-inter}, we obtain the general solution of structural system~\eqref{eq:structure-system}, given in case 6 of Table~\ref{tab1}.
	
	7. $k_{11}=k_{22}=1$, $k_{12}=k_{21}=0$, $m_a=0$.
We write systems~\eqref{eq:first-system-of-characteristic-system} and~\eqref{eq:second-system-of-characteristic-system} as
	\begin{gather}\label{eq:VII-class-for-uf}
		\dot u^1=u^1,\quad
		\dot u^2=u^2,\quad
		\dot f^{ab}=nf^{ab}.
	\end{gather}
The general solution of~\eqref{eq:VII-class-for-uf} coincides with~\eqref{eq:VI-class-gensol-inter} for $m=0$. Therefore the general solution of structural system~\eqref{eq:structure-system} is given in case 7 of Table~\ref{tab1}.

8. $k_{11}=k_{22}=k_{12}=1$, $k_{21}=0$, $m_a=0$.
Systems~\eqref{eq:first-system-of-characteristic-system} and~\eqref{eq:second-system-of-characteristic-system} take the form
\begin{gather}\label{eq:VIII-class-for-uf}
		\begin{aligned}
			\dot u^1	&=u^1+u^2,\quad
			\dot u^2=u^2,\\
			\dot f^{11}&=nf^{11}+f^{21},\quad
			\dot f^{12}=nf^{12}-f^{11}+f^{22},\quad
			\dot f^{21}=nf^{21},\quad
			\dot f^{22}=nf^{22}-f^{21}.
		\end{aligned}
	\end{gather}
The general solution of~\eqref{eq:VIII-class-for-uf} is
\begin{gather}\label{eq:VIII-class-gensol-inter}
		\begin{aligned}
			&u^1=(c_1+c_2p){\rm e}^p,\quad
			u^2=c_2{\rm e}^p,\\
			&f^{11}=\left(\lambda_{11}+\lambda_{21}p\right){\rm e}^{np},\quad
			f^{12}=\left(\lambda_{12}+\left(\lambda_{22}-\lambda_{11}\right)p-\lambda_{21}p^2\right){\rm e}^{np},\\
			&f^{21}=\lambda_{21}{\rm e}^{np},\quad
			f^{22}=\left(\lambda_{22}-\lambda_{21}p\right){\rm e}^{np},
		\end{aligned}
	\end{gather}
where $c_1$, $c_2$, and $\lambda_{ab}$ are constants of integration. Eliminating $p$ from~\eqref{eq:VIII-class-gensol-inter}, we obtain the general solution of structural system~\eqref{eq:structure-system}, given in case 8 of Table~\ref{tab1}.
	
	9. $k_{11}=k_{22}=m$,  $k_{12}=-k_{21}=1$, $m_a=0$.
Systems~\eqref{eq:first-system-of-characteristic-system} and~\eqref{eq:second-system-of-characteristic-system} are written as
	\begin{gather}\label{eq:IX-class-for-u}
		\begin{split}
		&\dot u^1=mu^1+u^2,\quad
		\dot u^2=-u^1+mu^2,\\
		&\dot f^{11}=nf^{11}+f^{12}+f^{21},\quad
		\dot f^{12}=nf^{12}-f^{11}+f^{22},\\
		&\dot f^{21}=nf^{21}-f^{11}+f^{22},\quad
		\dot f^{22}=nf^{22}-f^{12}-f^{21}.
		\end{split}
	\end{gather}
We first solve the first subsystem of~\eqref{eq:IX-class-for-u}.
To this end, we pass to polar coordinates
\[
u^1=\rho\cos\theta,\quad
u^2=\rho\sin\theta,
\]
where $\rho>0$ and $\theta=\arctan(u^2/u^1)$ on a local chart with $u^1>0$.
Substituting these expressions into the first subsystem of~\eqref{eq:IX-class-for-u}, we obtain
\[\dot\rho=m\rho, \quad
\dot\theta=-1.\]
Integrating this system gives
\begin{gather}\label{eq:IX-class-polar}
		\rho=\rho_0 {\rm e}^{mp},\quad
		\theta=\theta_0-p,
	\end{gather}
where $\rho_0>0$ and $\theta_0$ are constants of integration.

\noindent We take the following linear combinations of the equations for $f^{ab}$ in~\eqref{eq:IX-class-for-u}:
	\begin{gather}\label{eq:IX-class-linear-combination}
		\frac{\mathrm{ d}}{\mathrm{ d}p}\left(f^{11}+f^{22}\right)=n\left(f^{11}+f^{22}\right),\quad
		\frac{\mathrm{ d}}{\mathrm{ d}p}\left(f^{12}-f^{21}\right)=n\left(f^{12}-f^{21}\right).
	\end{gather}
From system~\eqref{eq:IX-class-linear-combination}, we find
\begin{gather}\label{eq:IX-class-inter}
		f^{22}=-f^{11}+2c_1{\rm e}^{np}, \quad 	
		f^{21}=f^{12}+2c_2{\rm e}^{np},
	\end{gather}
where $c_1$ and $c_2$ are constants of integration.
Substituting~\eqref{eq:IX-class-inter} into the third and fourth equations of~\eqref{eq:IX-class-for-u}, we obtain
	\begin{gather}\label{eq:IX-class-substitute}
		\dot f^{11}=nf^{11}+2f^{12}+2c_2{\rm e}^{np}, \quad 	
		\dot f^{12}=-2f^{11}+nf^{12}+2c_1{\rm e}^{np}.
	\end{gather}
The general solution of~\eqref{eq:IX-class-substitute} is
\begin{gather}\label{eq:IX-class-gensol-inter}
		f^{11}={\rm e}^{np}(c_3 \cos 2p+c_4 \sin 2p +c_1), \quad	
		f^{12}={\rm e}^{np}(-c_3 \sin 2p+c_4 \cos 2p -c_2),
	\end{gather}
where $c_1$, $c_2$, $c_3$, and $c_4$ are constants of integration.
Substituting~\eqref{eq:IX-class-gensol-inter} into~\eqref{eq:IX-class-inter}, we obtain
	\begin{gather}\label{eq:IX-class-gensol-inter2}
		f^{21}={\rm e}^{np}(-c_3 \sin 2p+c_4 \cos 2p +c_2), \quad	
		f^{22}={\rm e}^{np}(-c_3 \cos 2p-c_4 \sin 2p +c_1).
	\end{gather}
Thus the general solution of~\eqref{eq:IX-class-for-u} is given by~\eqref{eq:IX-class-polar},~\eqref{eq:IX-class-gensol-inter}, and~\eqref{eq:IX-class-gensol-inter2}.
Eliminating $p$ from these formulas, we obtain the general solution of structural system~\eqref{eq:structure-system}, listed in case 9 of Table~\ref{tab1}.
\end{proof}

\section{Group classification of diffusion systems from the first subclass}
		
In this section, we study the symmetry properties of systems belonging to the first subclass of class~\eqref{eq:mainsystem}.
These systems have the form
\begin{gather}\label{eq:I-class}
U_t=\p_x\left[{\rm e}^{nu^1}\varPhi\left(u^2\right)U_x\right],\quad
\varPhi:=\begin{pmatrix}
			\varphi^{11}&\varphi^{12}\\
			\varphi^{21}&\varphi^{22}
		\end{pmatrix},
\end{gather}
where $n\in\{0,1\}$ and
$\varphi^{ab}=\varphi^{ab}(u^2)$ are arbitrary smooth functions satisfying
$\varphi^{11}+\varphi^{22}\ne0$,
$\varphi^{11}\varphi^{22}-\varphi^{12}\varphi^{21}\ne0$.

Substituting the functions $f^{ab}$ from case 1 of Table~\ref{tab1} into the determining system~\eqref{eq:def-system} and splitting the resulting equations with respect to $u^1$, taking~\eqref{eq:coordinates-of-inf-operator} into account, we obtain the determining system for~\eqref{eq:I-class}:
		\begin{equation}\label{eq:I-class-define-system-n0}
			\begin{aligned}
				&\alpha^{21} \dot\varphi^{ab}=0,\quad
				 \alpha^{11}_x \dot \varphi^{a1}=0,\\
				&\left(\alpha^{22}u^2+\beta^2\right)\dot \varphi^{ab}+\left(\tau_t-2\xi_{x}\right)\varphi^{ab}+\alpha^{cb}\varphi^{ac}-\alpha^{ac}\varphi^{cb}=0,\\
&(\alpha^{12}_x u^2+\beta^1_x)\delta_{b2}\dot\varphi^{a1}
+(\alpha^{22}_x u^2+\beta^2_x)(\dot\varphi^{ab}+\delta_{b2}\dot\varphi^{a2})
+2\alpha^{cb}_x \varphi^{ac}-\xi_{xx} \varphi^{ab}+\delta_{ab}\xi_t=0,\\
				&\alpha^{a1}_t-\alpha^{b1}_{xx}\varphi^{ab}=0,\\
				&\alpha^{a2}_t u^2+\beta^a_t-\left(\alpha^{b2}_{xx} u^2+\beta^b_{xx}\right)\varphi^{ab}=0,
			\end{aligned}
		\end{equation}
for $n=0$, and
		\begin{equation}\label{eq:I-class-define-system-n1}
			\begin{aligned}
				&\alpha^{21}\dot\varphi^{ab}+\alpha^{11}\varphi^{ab}=0,\quad
				 \alpha^{11}_x\varphi^{a1}+\alpha^{21}_x\varphi^{a2}=0,\\
				&(\alpha^{22}u^2+\beta^2)\dot\varphi^{ab}+(\tau_t-2\xi_x+\alpha^{12}u^2+\beta^1)\varphi^{ab}
+\alpha^{cb}\varphi^{ac}-\alpha^{ac}\varphi^{cb}=0,\\
&(\alpha^{12}_x u^2+\beta^1_x)(\varphi^{ab}+\delta_{b1}\varphi^{a1}+\delta_{b2}\dot\varphi^{a1})
				+(\alpha^{22}_x u^2+\beta^2_x)(\dot\varphi^{ab}+\delta_{b1}\varphi^{a2}+\delta_{b2}\dot\varphi^{a2})
				\\&\qquad{}
				+2\alpha^{cb}_x \varphi^{ac}-\xi_{xx} \varphi^{ab}=0,
				\\
				&\left(\alpha^{b2}_{xx} u^2+\beta^b_{xx}\right)\varphi^{ab}=0,\\
				&\xi_t=\alpha^{ab}_t=\beta^a_t=0,
			\end{aligned}
		\end{equation}
for $n=1$,
where $\dot\varphi^{ab}=\mathrm{d}\varphi^{ab}/\mathrm{d}u^2$.

As stated in case 1 of Table~\ref{tab1}, the kernel of maximal Lie invariance algebras of systems from class~\eqref{eq:I-class} is the algebra
		\begin{gather}\label{eq:I-class-kernel}
			\bigl\langle\p_t, \, \p_x, \, D, \, nt\p_t-\p_{u^1}\bigr\rangle.
		\end{gather}
	
We determine the equivalence transformations of class~\eqref{eq:I-class}. The following theorem holds.
	\begin{theorem}\label{thm:equivalence-algebra-of-Iclass}
The equivalence algebra of class~\eqref{eq:I-class} is the linear span of the vector fields

		\begin{gather*}
\p_t,\ \p_x,\ \p_{u^1}-n\varphi^{ab}\p_{\varphi^{ab}},\ \p_{u^2},\\
t\p_t-\varphi^{ab}\p_{\varphi^{ab}},\quad x\p_x+2\varphi^{ab}\p_{\varphi^{ab}},\\
(1-n)(u^1\p_{u^1}+\varphi^{12}\p_{\varphi^{12}}-\varphi^{21}\p_{\varphi^{21}}),\\
u^2\p_{u^1}-nu^2\varphi^{ab}\p_{\varphi^{ab}}+\varphi^{21}(\p_{\varphi^{11}}-\p_{\varphi^{22}})
 +(\varphi^{22}-\varphi^{11})\p_{\varphi^{12}},\\
u^2\p_{u^2}-\varphi^{12}\p_{\varphi^{12}}+\varphi^{21}\p_{\varphi^{21}}
\end{gather*}
in the space $\mathbb R^2_{t,x}\times\mathbb R^2_U\times\mathbb R^4_{\varPhi}$.
	\end{theorem}
	\begin{proof}
Applying the algorithm for finding equivalence transformations proposed in~\cite{ahat-gaz-ibr}, we derive the determining system for the coefficients of the equivalence vector field
		\begin{gather}\label{eq:operator-of-equivalence-phi}
			E=\tau(t,x,U)\p_t+\xi(t,x,U)\p_x+\eta^a(t,x,U)\p_{u^a}+\zeta^{ab}(t,x,U,\varPhi)\p_{\varphi^{ab}}.
		\end{gather}
This system has the form
		\begin{gather}\label{eq:system-of-equivalence-phi}
			\begin{aligned}
				&\tau_{tt}=\tau_x=\tau_{u^a}=\xi_t=\xi_{xx}=\xi_{u^a}=\eta^a_t=\eta^a_x=n\eta^1_{u^1}=\eta^2_{u^1}=\eta^a_{u^bu^c}=0,\\
				&\zeta^{ab}=(2\xi_x-\tau_t-n\eta^1)\varphi^{ab}+\eta^a_{u^c}\varphi^{cb}-\eta^c_{u^b}\varphi^{ac}.
			\end{aligned}
		\end{gather}
The general solution of system~\eqref{eq:system-of-equivalence-phi} is
		\begin{gather}\label{eq:gensol-of-equivalence-phi}
			\begin{split}
			&\tau=c_1t+d_0,\quad
			\xi=c_2x+d_1,\quad
			\eta^1=k_{11}u^1+k_{12}u^2+m_1,\quad
			\eta^2=k_{22}u^2+m_2,\\
			&\zeta^{ab}=(2c_2-c_1-n(k_{12}u^2+m_1))\varphi^{ab}+k_{1c}(\delta_{a1}\varphi^{cb}-\delta_{bc}\varphi^{a1})
+k_{22}(\delta_{a2}\varphi^{2b}-\delta_{b2}\varphi^{a2}).
			\end{split}
		\end{gather}
where $c_1$, $c_2$, $d_0$, $d_1$, $k_{ab}$, and $m_a$ are arbitrary constants with $nk_{11}=k_{21}=0$.
Substituting~\eqref{eq:gensol-of-equivalence-phi} into~\eqref{eq:operator-of-equivalence-phi}, we collect the coefficients of the arbitrary constants.
This yields the basis vector fields stated in the theorem.
	\end{proof}

\begin{corollary}
The group of continuous equivalence transformations of class~\eqref{eq:I-class}
consists of the point transformations
\begin{gather*}
\tilde t={\rm e}^{\theta_1}t+s_0,\quad
\tilde x={\rm e}^{\theta_2}x+s_1,\quad
\tilde U=AU+Q,\quad
\tilde\varPhi={\rm e}^{2\theta_2-\theta_1-nq_1-n\theta_5 u^2}A\varPhi A^{-1},
\end{gather*}
where $s_0$, $s_1$, $q_1$, $q_2$, and $\theta_1,\dots,\theta_5$ are arbitrary constants,
\[
A=\begin{pmatrix}
{\rm e}^{\theta_3} & \theta_5\\
0 & {\rm e}^{\theta_4}
\end{pmatrix}\ \ \text{for }n=0,\quad
A=\begin{pmatrix}
		1 & \theta_5\\
		0 & {\rm e}^{\theta_4}
\end{pmatrix}\ \ \text{for }n=1,\quad
Q=\begin{pmatrix}q_1\\q_2\end{pmatrix}.
\]
\end{corollary}

\begin{remark}
In addition to continuous transformations,
class~\eqref{eq:I-class} admits the following discrete equivalence transformations:
\begin{gather*}
		\tilde t=-t, \quad \tilde x=x, \quad \tilde u^1=u^1,
		\quad \tilde u^2=u^2, \quad \tilde \varphi^{ab}=-\varphi^{ab},\\
		\tilde t=t, \quad \tilde x=-x, \quad \tilde u^1=u^1,
		\quad \tilde u^2=u^2, \quad \tilde \varphi^{ab}=\varphi^{ab},\\
		\tilde t=t, \quad \tilde x=x, \quad \tilde u^1=u^1,
		\quad \tilde u^2=-u^2, \quad \tilde \varphi^{11}=\varphi^{11}, \quad \tilde \varphi^{12}=-\varphi^{12},\\
\tilde \varphi^{21}=-\varphi^{21}, \quad \tilde \varphi^{22}=\varphi^{22}
	\end{gather*}
and, in addition, for $n=0$,
	\begin{gather*}
\tilde t=t, \quad \tilde x=x, \quad \tilde u^1=-u^1, \quad \tilde u^2=u^2,\\
\tilde \varphi^{11}=\varphi^{11}, \quad \tilde \varphi^{12}=-\varphi^{12}, \quad
\tilde \varphi^{21}=-\varphi^{21}, \quad \tilde \varphi^{22}=\varphi^{22}.
	\end{gather*}
These continuous and discrete transformations generate the following equivalence transformations of class~\eqref{eq:I-class}:

	\begin{gather}\label{eq:disc-transform-of-equivalence-Iclass-super}
	\tilde t=l_0t+s_0, \quad \tilde x=l_1x+s_1, \quad \tilde U=PU+Q, \quad
\tilde\varPhi=\frac{l_1^2}{l_0}{\rm e}^{-np_{12} u^2-nq_1}P\varPhi P^{-1},
	\end{gather}
where $l_0$, $l_1$, $p_{11}$, $p_{12}$, and $p_{22}$ are arbitrary constants satisfying $l_0l_1p_{11}p_{22}\ne0$,
\[
P:=\begin{pmatrix}
	p_{11} & p_{12}\\
	0 & p_{22}
\end{pmatrix}\ \ \text{for }n=0,\quad
P:=\begin{pmatrix}
		1 & p_{12}\\
		0 & p_{22}
\end{pmatrix}\ \ \text{for }n=1,\quad
Q=\begin{pmatrix}q_1\\q_2\end{pmatrix}.
\]
\end{remark}

We carry out the group classification of class~\eqref{eq:I-class} up to the equivalence transformations found above. That is, we determine the forms of arbitrary elements $\varphi^{ab}$ for which a system from class~\eqref{eq:I-class} admits an invariance algebra larger than~\eqref{eq:I-class-kernel}.

\begin{theorem}\label{thm:for-table-of-the-I-class}
The complete group classification of systems of the first class~\eqref{eq:I-class},
up to equivalence transformations~\eqref{eq:disc-transform-of-equivalence-Iclass-super},
is given in Table~\ref{tab2} for $n=0$ and Table~\ref{tab3} for $n=1$;
case~0 in each table gives the kernel of maximal Lie invariance algebras of the corresponding subclass.
\end{theorem}

\noindent In Tables~\ref{tab2} and~\ref{tab3}, $G_{1,n}^\sim$ denotes the full
equivalence group of the first class for the corresponding value of $n$; its transformations
are given in~\eqref{eq:disc-transform-of-equivalence-Iclass-super}.
Conditions labeled $\bmod G_{1,n}^\sim$ specify a choice of canonical
representatives, not restrictions on the original system. A dot denotes
differentiation with respect to $u^2$. By convention, general rows do not repeat exclusions corresponding
to separate cases listed elsewhere in the same table.

\begingroup
\footnotesize
\setlength{\tabcolsep}{4pt}
\begin{longtable}{|>{\raggedright\arraybackslash}p{5mm}|>{\raggedright\arraybackslash}p{58mm}|>{\raggedright\arraybackslash}p{54mm}|>{\raggedright\arraybackslash}p{30mm}|}
	\captionsetup{width=\textwidth}
\caption{Group classification of diffusion systems of the first class for $n=0$} \label{tab2} \\
	\hline
 No. & $\varphi^{ab}$ & Basis of $A^{\max}$ & Conditions \\
	\hline
	\endfirsthead
	\hline
 No. & $\varphi^{ab}$ & Basis of $A^{\max}$ & Conditions \\
	\hline
	\endhead
	\hline
	\endfoot
	0&$\forall$  &$A^{\cap}_1=\langle A^{\cap}, \ \p_{u^1}\rangle$&$\varphi^{22}\ne-\varphi^{11}$, \\
	&&&$\varphi^{11}\varphi^{22}\ne\varphi^{12}\varphi^{21}$\\
	\hline
	1 & $\varphi^{11}=1$, $\varphi^{12}=0$,
& $AG_2(1,1)$ with $Q_1:=u^1\p_{u^1}+\lambda^{-1}u^2\p_{u^2}$,
	& $\lambda\ne0$, $\pm1$,
	\\
	& $\varphi^{21}=0$, $\varphi^{22}=\lambda$
	& $u^1\p_{u^1}, \, \beta^a(t,x)\p_{u^a}$
	& $\beta^1_t=\beta^1_{xx}$,  \\
	&&&$\beta^2_t=\lambda\beta^2_{xx}$\\
	\hline
	
 2 & $\varphi^{11}=1$, $\varphi^{12}=1$, & $AG_2(1,1)$ with $Q_1:=I-u^2\p_{u^1}$, &$\beta^1_t=\beta^1_{xx}+\beta^2_{xx}$,\\
	&$\varphi^{21}=0$, $\varphi^{22}=1$   & $u^2\p_{u^1}, \, \beta^a(t,x)\p_{u^a}$ &$\beta^2_t=\beta^2_{xx}$
	\\
	\hline
	
 3 &$\varphi^{11}=1$, $\varphi^{12}=0$, & $AG_2(1,1)$ with $Q_1:=I$, &$\beta^a_t=\beta^a_{xx}$\\
	&$\varphi^{21}=0$, $\varphi^{22}=1$  &\shortstack[l]{$u^1\p_{u^1}, \, u^2\p_{u^1},$\\$u^1\p_{u^2}, \, \beta^a(t,x)\p_{u^a}$}  &\\
	\hline
	
	4&$\varphi^{11}=1,\quad \varphi^{12}=\eta$,
&$AG_2(1,1)$ with $Q_1:=\dfrac{(u^1-\eta u^2)\p_{u^1}+(u^2-u^1)\p_{u^2}}{1-\eta}$,
	&$\eta\in\mathbb R\setminus\{1\}$,\\
	&$\varphi^{21}=1,\quad\varphi^{22}=1$
	&$J_\eta:=\eta u^2\p_{u^1}+u^1\p_{u^2},$
	&$\beta^1_t=\beta^1_{xx}+\eta\beta^2_{xx}$,\\
	&&$\beta^a(t,x)\p_{u^a}$
	&$\beta^2_t=\beta^1_{xx}+\beta^2_{xx}$\\
	\hline
	5&$\varphi^{11}=0, \ \varphi^{12}\text{ arbitrary}$,   &$A^{\cap}_1, \ x\p_{u^1}$&$\varphi^{22}\ne0$,\\
	& $\varphi^{21}=\lambda_{21}, \ \varphi^{22}\text{ arbitrary}$&&$\lambda_{21}\varphi^{12}\ne0$\\
	&&&$(\dot\varphi^{12},\dot\varphi^{22})\not\equiv(0,0)$;\\
	&&&$\lambda_{21}=1\bmod G_{1,0}^\sim$\\

	\hline
	6&$\varphi^{11}=\lambda_{11}, \ \varphi^{12}\text{ arbitrary}$,  &$A^{\cap}, \ \beta^1(t,x)\p_{u^1}$&$\varphi^{22}\ne-\lambda_{11}$,\\
	&$\varphi^{21}=0, \ \varphi^{22}\text{ arbitrary}$&&$\lambda_{11}\varphi^{22}\ne0$,\\
	&&&$\beta^1_t=\lambda_{11}\beta^1_{xx}$\\
	&&&$(\dot\varphi^{12},\dot\varphi^{22})\not\equiv(0,0)$;\\
	&&&$\lambda_{11}=1\bmod G_{1,0}^\sim$\\

	\hline
	7&$\varphi^{11}\text{ arbitrary}, \ \varphi^{12}=0$,   &$A^{\cap}_1, \ u^1\p_{u^1}, \ u^2\p_{u^1}$&$\varphi^{11}\ne0$\\
	&$\varphi^{21}=0, \ \varphi^{22}\text{ arbitrary}, \ (\varphi^{11}=\varphi^{22})$&&\\
	&&&$\dot\varphi^{11}\not\equiv0$\\

	\hline
	8&$\varphi^{11}=\lambda_{11}(u^2)^p, \ \varphi^{12}=0$,&$A^{\cap}_1, \ px\p_x+2I, \ u^1\p_{u^1}, \ u^2\p_{u^1}$&$\lambda_{11}\ne0$\\
	&$\varphi^{21}=0, \ \varphi^{22}=\lambda_{11}(u^2)^p$&&\\
		&&&$p\notin\{0,-2\}$;\\
	&&&$\lambda_{11}=1\bmod G_{1,0}^\sim$\\

	\hline
	9&$\varphi^{11}=\lambda_{11}{\rm e}^{pu^2}, \ \varphi^{12}=0$, &$A^{\cap}_1, \ px\p_x+2\p_{u^2}, \ u^1\p_{u^1}, \ u^2\p_{u^1}$&$\lambda_{11}\ne0$\\
	&$\varphi^{21}=0, \ \varphi^{22}=\lambda_{11}{\rm e}^{pu^2}$&&\\
		&&&$p\ne0$;\\
	&&&$\lambda_{11}=p=1\bmod G_{1,0}^\sim$\\

	\hline
	10&$\varphi^{11}=\lambda_{11}(u^2)^{-2}, \ \varphi^{12}=0$,&$A^{\cap}_1, \ -x\p_x+I, \ u^1\p_{u^1}, \ u^2\p_{u^1},$&$\lambda_{11}\ne0$;\\
	&$\varphi^{21}=0, \ \varphi^{22}=\lambda_{11}(u^2)^{-2}$&$xu^2\p_{u^1}$&\\
	&&&$\lambda_{11}=1\bmod G_{1,0}^\sim$\\

	\hline
	11&$\varphi^{11}\text{ arbitrary}, \ \varphi^{12}=0$,   &$A^{\cap}_1, \ u^1\p_{u^1}$&$\varphi^{22}\ne-\varphi^{11}$,\\
	&$\varphi^{21}=0, \ \varphi^{22}\text{ arbitrary}, \ (\varphi^{11}\ne\varphi^{22})$&&$\varphi^{11}\varphi^{22}\ne0$\\
	&&&$\dot\varphi^{11}\not\equiv0$\\

	\hline
	12&$\varphi^{11}=\lambda_{11}(u^2)^p, \ \varphi^{12}=0$,&$A^{\cap}_1, \ px\p_x+2I, \ u^1\p_{u^1}$&$\lambda_{22}\ne-\lambda_{11}$,\\
	&$\varphi^{21}=0, \ \varphi^{22}=\lambda_{22}(u^2)^p$&&$\lambda_{11}\lambda_{22}\ne0$\\
		&&&$p\ne0,$\\
	&&&$\lambda_{22}\ne\lambda_{11}$;\\
	&&&$\lambda_{11}=1\bmod G_{1,0}^\sim$\\

	\hline
	13&$\varphi^{11}=\lambda_{11}{\rm e}^{pu^2}, \ \varphi^{12}=0$, &$A^{\cap}_1, \ px\p_x+2\p_{u^2}, \ u^1\p_{u^1}$&$\lambda_{22}\ne-\lambda_{11}$,\\
	&$\varphi^{21}=0, \ \varphi^{22}=\lambda_{22}{\rm e}^{pu^2}$&&$\lambda_{11}\lambda_{22}\ne0$\\
		&&&$p\ne0,$\\
	&&&$\lambda_{22}\ne\lambda_{11}$;\\
	&&&$\lambda_{11}=p=1\bmod G_{1,0}^\sim$\\

	\hline
	14&$\varphi^{11}=\lambda_{11}, \ \varphi^{12}=0$,   &$A^{\cap}, \ \beta^1(t,x)\p_{u^1}, \ u^1\p_{u^1}$&$\varphi^{22}\ne-\lambda_{11}$,\\
	&$\varphi^{21}=0, \ \varphi^{22}\text{ arbitrary}$&&$\lambda_{11}\varphi^{22}\ne0$,\\
	&&&$\beta^1_t=\lambda_{11}\beta^1_{xx}$\\
	&&&$\dot\varphi^{22}\not\equiv0$;\\
	&&&$\lambda_{11}=1\bmod G_{1,0}^\sim$\\

	\hline
	15&$\varphi^{11}\text{ arbitrary}, \ \varphi^{12}\text{ arbitrary}$,   &$A^{\cap}_1, \ u^2\p_{u^1}$&$\varphi^{11}\ne0$,\\
	&&&$\varphi^{12}\not\equiv0$\\
	&$\varphi^{21}=0, \ \varphi^{22}\text{ arbitrary}, \ (\varphi^{11}=\varphi^{22})$&&\\
	&&&$\dot\varphi^{11}\not\equiv0$\\

	\hline
	16&$\varphi^{11}=\lambda_{11}(u^2)^p, \ \varphi^{12}=\lambda_{12}(u^2)^{p-s}$,&$A^{\cap}_1, \ px\p_x+2(I-su^1\p_{u^1})$,&$\lambda_{11}\ne0$\\
	&$\varphi^{21}=0, \ \varphi^{22}=\lambda_{11}(u^2)^p$&$u^2\p_{u^1}$&\\
		&&&$p\notin\{0,-2\},$\\
	&&&$\lambda_{12}\ne0$;\\
	&&&$\lambda_{11}=\lambda_{12}=1\bmod G_{1,0}^\sim$\\

	\hline
	17&$\varphi^{11}=\lambda_{11}{\rm e}^{pu^2}, \ \varphi^{12}=\lambda_{12}{\rm e}^{(p+s)u^2}$, &$A^{\cap}_1, \ px\p_x+2su^1\p_{u^1}+2\p_{u^2}$,&$\lambda_{11}\ne0$\\
	&$\varphi^{21}=0, \ \varphi^{22}=\lambda_{11}{\rm e}^{pu^2}$&$u^2\p_{u^1}$&\\
		&&&$p\ne0,$\\
	&&&$\lambda_{12}\ne0$;\\
	&&&$\lambda_{11}=\lambda_{12}=p=1\bmod G_{1,0}^\sim$\\

	\hline
	18&$\varphi^{11}=\lambda_{11}, \ \varphi^{12}\text{ arbitrary}$,   &$A^{\cap}, \ \beta^1(t,x)\p_{u^1}, \ u^2\p_{u^1}$,&$\lambda_{11}\ne0$,\\
	&$\varphi^{21}=0, \ \varphi^{22}=\lambda_{11}$&&$\beta^1_t=\lambda_{11}\beta^1_{xx}$\\
	&&&$\dot\varphi^{12}\not\equiv0$;\\
	&&&$\lambda_{11}=1\bmod G_{1,0}^\sim$\\

	\hline
	19&$\varphi^{11}=\lambda_{11}, \ \varphi^{12}=\lambda_{12}(u^2)^{-s}$,  &$A^{\cap}, \ \beta^1(t,x)\p_{u^1}$,&$\lambda_{11}\ne0$,\\
	&$\varphi^{21}=0, \ \varphi^{22}=\lambda_{11}$ &$I-su^1\p_{u^1}, \ u^2\p_{u^1}$&$\beta^1_t=\lambda_{11}\beta^1_{xx}$\\
		&&&$s\ne0,\ \lambda_{12}\ne0$;\\
	&&&$\lambda_{11}=\lambda_{12}=1\bmod G_{1,0}^\sim$\\

	\hline
	20&$\varphi^{11}=\lambda_{11}, \ \varphi^{12}=\lambda_{12}{\rm e}^{su^2}$,   &$A^{\cap}, \ \beta^1(t,x)\p_{u^1}$,&$\lambda_{11}\ne0$,\\
	&$\varphi^{21}=0, \ \varphi^{22}=\lambda_{11}$ &$su^1\p_{u^1}+\p_{u^2}, \ u^2\p_{u^1}$&$\beta^1_t=\lambda_{11}\beta^1_{xx}$\\
		&&&$s\ne0,\ \lambda_{12}\ne0$;\\
	&&&$\lambda_{11}=\lambda_{12}=s=1\bmod G_{1,0}^\sim$\\

	\hline
	21&$\varphi^{11}=\lambda_{11}(u^2)^{-2}, \ \varphi^{12}\text{ arbitrary}$,&$A^{\cap}_1,  \ u^2\p_{u^1}, \ xu^2\p_{u^1}$&$\lambda_{11}\ne0$;\\
	&$\varphi^{21}=0, \ \varphi^{22}=\lambda_{11}(u^2)^{-2}$&&\\
	&&&$\lambda_{11}=1\bmod G_{1,0}^\sim$\\

	\hline
	22&$\varphi^{11}=\lambda_{11}(u^2)^{-2}, \ \varphi^{12}=\lambda_{12}(u^2)^{-2-s}$,&$A^{\cap}_1, \ x\p_x+su^1\p_{u^1}-I$,&$\lambda_{11}\ne0$\\
	&$\varphi^{21}=0, \ \varphi^{22}=\lambda_{11}(u^2)^{-2}$&$u^2\p_{u^1}, \ xu^2\p_{u^1}$&\\
		&&&$\lambda_{12}\ne0$;\\
	&&&$\lambda_{11}=\lambda_{12}=1\bmod G_{1,0}^\sim$\\

	\hline
	23&$\varphi^{11}=\lambda_{11}(u^2)^p, \ \varphi^{12}=\lambda_{12}(u^2)^{p-s}$, &$A^{\cap}_1, \ px\p_x+2(I-su^1\p_{u^1})$&$\lambda_{22}\ne-\lambda_{11}$,\\
	&$\varphi^{21}=\lambda_{21}(u^2)^{p+s}, \ \varphi^{22}=\lambda_{22}(u^2)^p$ &&$\lambda_{11}\lambda_{22}\ne\lambda_{21}\lambda_{12}$\\
		&&&$(p,s\lambda_{12},$\\
	&&&$s\lambda_{21})\ne(0,0,0)$,\\

	&&&$(p,\lambda_{21})\ne(0,0)$\\
	&&&$(\lambda_{12},\lambda_{21})\ne(0,0)$\\
	&&&$(s,\lambda_{21})\ne(0,0)$,\\
	&&&$(\lambda_{21},\lambda_{22}-\lambda_{11})\ne(0,0)$,\\
	&&&$(\lambda_{11},p+s)\ne(0,0)$;\\
	&&&$\lambda_{11}+\lambda_{22}=1\bmod G_{1,0}^\sim$\\
	&&&$\lambda_{12}=1\ (\lambda_{12}\ne0),$\\
	&&&$\lambda_{21}=1\ (\lambda_{12}=0)\bmod G_{1,0}^\sim$\\
	&&&$\lambda_{11}=\lambda_{22}=\tfrac12$\\
	&&&(if $s=0$) $\bmod G_{1,0}^\sim$\\

	\hline
	24&$\varphi^{11}=\lambda_{11}, \ \varphi^{12}=\lambda_{12}(u^2)^{-s}$,  &$A^{\cap}, \ \beta^1(t,x)\p_{u^1}, \ I-su^1\p_{u^1}$&$\lambda_{22}\ne-\lambda_{11}$,\\
	&$\varphi^{21}=0, \ \varphi^{22}=\lambda_{22}$ &&$\lambda_{11}\lambda_{22}\ne0$,\\
	&&&$\beta^1_t=\lambda_{11}\beta^1_{xx}$\\
		&&&$s\notin\{0,-1\},$\\
	&&&$\lambda_{12}\ne0,$\\
	&&&$\lambda_{22}\ne\lambda_{11}$;\\
	&&&$\lambda_{11}=\lambda_{12}=1\bmod G_{1,0}^\sim$\\

	\hline
	25&$\varphi^{11}=\lambda_{11}, \ \varphi^{12}=\lambda_{12}u^2$,  &$A^{\cap}, \ \beta^1(t,x)\p_{u^1}, \ 2u^1\p_{u^1}+u^2\p_{u^2}$,&$\lambda_{22}\ne-\lambda_{11}$,\\
	&$\varphi^{21}=0, \ \varphi^{22}=\lambda_{22}$ &$\lambda_{12}u^2\p_{u^1}+(\lambda_{22}-\lambda_{11})\p_{u^2}$&$\lambda_{11}\lambda_{22}\ne0$,\\
	&&&$\beta^1_t=\lambda_{11}\beta^1_{xx}$\\
		&&&$\lambda_{12}\ne0,$\\
	&&&$\lambda_{22}\ne\lambda_{11}$;\\
	&&&$\lambda_{11}=\lambda_{12}=1\bmod G_{1,0}^\sim$\\

	\hline
	26&$\varphi^{11}=0, \ \varphi^{12}=\lambda_{12}(u^2)^{2p}$,  &$A^{\cap}_1, \ px\p_x+2(pu^1\p_{u^1}+I), \ x\p_{u^1}$&$\lambda_{22}\ne0$,\\
	&$\varphi^{21}=\lambda_{21}, \ \varphi^{22}=\lambda_{22}(u^2)^p$ &&$\lambda_{12}\lambda_{21}\ne0$\\
		&&&$p\ne0$;\\
	&&&$\lambda_{22}=\lambda_{12}=1\bmod G_{1,0}^\sim$\\

	\hline
	27
	&$\varphi^{11}=(\lambda_{11}+a\lambda_{21}\ln u^2)(u^2)^p$,  &$A^{\cap}_1, \ px\p_x+2(I+au^2\p_{u^1})$&$\lambda_{22}\ne-\lambda_{11}$,
	\\
	&$\varphi^{12}=(\lambda_{12}+a(\lambda_{22}-\lambda_{11})\ln u^2$&&$\lambda_{11}\lambda_{22}\ne\lambda_{21}\lambda_{12}$\\
	&$\qquad -a^2\lambda_{21}\ln^2 u^2)(u^2)^p$,&&
	\\
	&$\varphi^{21}=\lambda_{21}(u^2)^p$,&&\\
	&$\varphi^{22}=(\lambda_{22}-a\lambda_{21}\ln u^2)(u^2)^p$&&\\
		&&&$a\ne0,$\\
	&&&$(\lambda_{21},\lambda_{22}-\lambda_{11})$\\
	&&&$\ne(0,0)$,\\
	&&&$(p,\lambda_{21})\ne(0,0)$;\\
	&&&$\lambda_{11}+\lambda_{22}=a=1\bmod G_{1,0}^\sim$\\
	&&&$\lambda_{11}=\lambda_{22}=\tfrac12\ (\lambda_{21}\ne0),$\\
	&&&$\lambda_{12}=0\ (\lambda_{21}=0)\bmod G_{1,0}^\sim$\\

	\hline
	28&$\varphi^{11}=\lambda_{11}$,  &$A^{\cap}, \ \beta^1(t,x)\p_{u^1}, \ I+au^2\p_{u^1},$&$\lambda_{22}\ne-\lambda_{11}$,\\
	&$\varphi^{12}=\lambda_{12}+a(\lambda_{22}-\lambda_{11})\ln u^2$, &&$\lambda_{11}\lambda_{22}\ne0$,\\
	&$\varphi^{21}=0, \ \varphi^{22}=\lambda_{22}$&&$\beta^1_t=\lambda_{11}\beta^1_{xx}$\\
		&&&$a\ne0,$\\
	&&&$\lambda_{22}\ne\lambda_{11}$;\\
	&&&$\lambda_{11}=a=1,\ \lambda_{12}=0$\\
	&&&$\bmod G_{1,0}^\sim$\\

	\hline
	29&$\varphi^{11}=\lambda_{11}{\rm e}^{pu^2}, \ \varphi^{12}=\lambda_{12}{\rm e}^{(p+s)u^2}$,  &$A^{\cap}_1, \ px\p_x+2su^1\p_{u^1}+2\p_{u^2}$&$\lambda_{22}\ne-\lambda_{11}$,\\
	&$\varphi^{21}=\lambda_{21}{\rm e}^{(p-s)u^2}, \ \varphi^{22}=\lambda_{22}{\rm e}^{pu^2}$&&$\lambda_{11}\lambda_{22}\ne\lambda_{21}\lambda_{12}$\\
		&&&$(p,s\lambda_{12},$\\
	&&&$s\lambda_{21})\ne(0,0,0)$,\\

	&&&$p\ne0,\ (\lambda_{12},\lambda_{21})\ne(0,0)$\\
	&&&$(s,\lambda_{21})\ne(0,0)$\\
	&&&$(\lambda_{21},\lambda_{22}-\lambda_{11})\ne(0,0)$,\\
	&&&$(\lambda_{11},p-s)\ne(0,0)$;\\
	&&&$p=\lambda_{11}+\lambda_{22}=1\bmod G_{1,0}^\sim$\\
	&&&$\lambda_{12}=1\ (\lambda_{12}\ne0),$\\
	&&&$\lambda_{21}=1\ (\lambda_{12}=0)\bmod G_{1,0}^\sim$\\
	&&&$\lambda_{11}=\lambda_{22}=\tfrac12$\\
	&&&(if $s=0$) $\bmod G_{1,0}^\sim$\\

	\hline
	30&$\varphi^{11}=0, \ \varphi^{12}=\lambda_{12}{\rm e}^{2pu^2}$,  &$A^{\cap}_1, \ px\p_x+2pu^1\p_{u^1}+2\p_{u^2}, \ x\p_{u^1}$&$\lambda_{22}\ne0$,\\
	&$\varphi^{21}=\lambda_{21}, \ \varphi^{22}=\lambda_{22}{\rm e}^{pu^2}$&&$\lambda_{12}\lambda_{21}\ne0$\\
		&&&$p\ne0$;\\
	&&&$\lambda_{22}=\lambda_{12}=p=1\bmod G_{1,0}^\sim$\\

	\hline
	31
	&$\varphi^{11}=(\lambda_{11}+a\lambda_{21}u^2){\rm e}^{pu^2}$,  &$A^{\cap}_1, \ px\p_x+2au^2\p_{u^1}+2\p_{u^2}$&$\lambda_{22}\ne-\lambda_{11}$,
	\\
	&$\varphi^{12}=(\lambda_{12}+a(\lambda_{22}-\lambda_{11})u^2$&&$\lambda_{11}\lambda_{22}\ne\lambda_{21}\lambda_{12}$\\
	&$\qquad -a^2\lambda_{21}(u^2)^2){\rm e}^{pu^2}$,&&
	\\
	&$\varphi^{21}=\lambda_{21}{\rm e}^{pu^2}$,&&\\
	&$\varphi^{22}=(\lambda_{22}-a\lambda_{21}u^2){\rm e}^{pu^2}$&&\\
		&&&$p\ne0,\ a\ne0,$\\
	&&&$(\lambda_{21},\lambda_{22}-\lambda_{11})$\\
	&&&$\ne(0,0)$;\\
	&&&$p=a=\lambda_{11}+\lambda_{22}=1$\\
	&&&$\bmod G_{1,0}^\sim$\\
	&&&$\lambda_{11}=\lambda_{22}=\tfrac12\ (\lambda_{21}\ne0),$\\
	&&&$\lambda_{12}=0\ (\lambda_{21}=0)\bmod G_{1,0}^\sim$\\

	\hline
	32&$\varphi^{11}=\lambda_{11}, \ \varphi^{12}=\lambda_{12}{\rm e}^{su^2}$,   &$A^{\cap}_1, \ su^1\p_{u^1}+\p_{u^2}$&$\lambda_{22}\ne-\lambda_{11}$,\\
	&$\varphi^{21}=\lambda_{21}{\rm e}^{-su^2}, \ \varphi^{22}=\lambda_{22}$ &&$\lambda_{11}\lambda_{22}\ne\lambda_{21}\lambda_{12}$\\
		&&&$s\ne0,\ \lambda_{21}\ne0$;\\
	&&&$s=\lambda_{21}=\lambda_{11}+\lambda_{22}=1$\\
	&&&$\bmod G_{1,0}^\sim$\\

	\hline
	33&$\varphi^{11}=\lambda_{11}, \ \varphi^{12}=\lambda_{12}{\rm e}^{su^2}$,   &$A^{\cap}, \ \beta^1(t,x)\p_{u^1}, \ su^1\p_{u^1}+\p_{u^2}$&$\lambda_{22}\ne-\lambda_{11}$,\\
	&$\varphi^{21}=0, \ \varphi^{22}=\lambda_{22}$ &&$\lambda_{11}\lambda_{22}\ne0$,\\
	&&&$\beta^1_t=\lambda_{11}\beta^1_{xx}$\\
		&&&$s\ne0,\ \lambda_{12}\ne0,$\\
	&&&$\lambda_{22}\ne\lambda_{11}$;\\
	&&&$s=\lambda_{11}=\lambda_{12}=1\bmod G_{1,0}^\sim$\\

	\hline

	34&$\varphi^{11}=1+u^2,\quad\varphi^{12}=\eta-(u^2)^2$, & $A^{\cap}_1,\ u^2\p_{u^1}+\p_{u^2}$&$\eta\ne1$\\
	&$\varphi^{21}=1,\qquad\varphi^{22}=1-u^2$&&\\

	\hline
\end{longtable}

$AG_2(1,1)$ denotes the generalized Galilei algebra with basis
\begin{gather*}
	\p_t,\ D_{\mathrm G}:=2t\p_t+x\p_x-\tfrac12I,\ \Pi:=t^2\p_t+tx\p_x-\tfrac14x^2Q_1-\tfrac12tI,\
	\p_x,\ G:=t\p_x-\tfrac12xQ_1,\ Q_1,
\end{gather*}
which
is isomorphic to ${\rm sl}(2,\mathbb R)\lsemioplus_{\rho_1\oplus \rho_0}{\rm h}(1,\mathbb R)$~\cite[Section~2]{kova2023b},
with $I:=u^a\p_{u^a}$, $J:=u^2\p_{u^1}-u^1\p_{u^2}$, and $Q_1$ specified in the corresponding case.

\begin{longtable}{|>{\raggedright\arraybackslash}p{5mm}|>{\raggedright\arraybackslash}p{58mm}|>{\raggedright\arraybackslash}p{54mm}|>{\raggedright\arraybackslash}p{30mm}|}
	\captionsetup{width=\textwidth}
\caption{Group classification of diffusion systems of the first class for $n=1$} \label{tab3} \\
	\hline
 No. & $\varphi^{ab}$ & Basis of $A^{\max}$ & Conditions \\
	\hline
	\endfirsthead
	\hline
 No. & $\varphi^{ab}$ & Basis of $A^{\max}$ & Conditions \\
	\hline
	\endhead
	\hline
	\endfoot
	0&$\forall$  &$A^{\cap}_1=\langle A^{\cap}, \ t\p_t-\p_{u^1}\rangle$&$\varphi^{22}\ne-\varphi^{11}$, \\
	&&&$\varphi^{11}\varphi^{22}\ne\varphi^{12}\varphi^{21}$\\
	\hline
	1&$\varphi^{11}=\lambda_{11}, \ \varphi^{12}=\lambda_{12}$, &$A^{\cap}_1, \ \p_{u^2}$&$\lambda_{22}\ne-\lambda_{11}$,\\
	&$\varphi^{21}=\lambda_{21}, \ \varphi^{22}=\lambda_{22}$&&$\lambda_{11}\lambda_{22}\ne\lambda_{12}\lambda_{21}$\\
	&&&$\lambda_{12}\ne0$;\\
	&&&$\lambda_{12}=1$,\\
	&&&$\lambda_{11}+\lambda_{22}=1\bmod G_{1,1}^\sim$\\

	\hline
	2&$\varphi^{11}=\lambda_{11},\ \varphi^{12}=0$, &$A^{\cap}_1,\ \p_{u^2},\ \lambda_{21}u^1\p_{u^2}+(\lambda_{22}-\lambda_{11})u^2\p_{u^2}$&$\lambda_{22}\ne-\lambda_{11}$,\\
	&$\varphi^{21}=\lambda_{21},\ \varphi^{22}=\lambda_{22}$&&$\lambda_{11}\lambda_{22}\ne0$,\\
	&&&$(\lambda_{21},\lambda_{22}-\lambda_{11})\ne(0,0)$;\\
	&&&$\lambda_{11}=1\bmod G_{1,1}^\sim$\\
	&&&$\lambda_{21}=1\ (\lambda_{21}\ne0)\bmod G_{1,1}^\sim$\\

	\hline
	3&$\varphi^{11}=\lambda_{22}, \ \varphi^{12}=0$, &$A^{\cap}_1, \ \p_{u^2}, \ u^1\p_{u^2}, \ u^2\p_{u^2}$&$\lambda_{22}\ne0$;\\
	&$\varphi^{21}=0, \ \varphi^{22}=\lambda_{22}$&&\\
	&&&$\lambda_{22}=1\bmod G_{1,1}^\sim$\\

	\hline

4 & $\varphi^{11}=(\lambda_{11}-\varepsilon\lambda_{21}u^2)\mathrm e^{\varepsilon(u^2)^2/2}$,
 & $A^{\cap}_1,\ Q_\varepsilon$
 & $\varepsilon=\pm1$,\\
 & $\varphi^{12}=\bigl(\lambda_{12}-\varepsilon(\lambda_{22}-\lambda_{11})u^2$ && $\lambda_{22}\ne-\lambda_{11}$,\\
 & $\qquad-\lambda_{21}(u^2)^2\bigr)\mathrm e^{\varepsilon(u^2)^2/2}$, && $\lambda_{11}\lambda_{22}\ne\lambda_{12}\lambda_{21}$,\\
 & $\varphi^{21}=\lambda_{21}\mathrm e^{\varepsilon(u^2)^2/2}$, && $(\lambda_{22},\lambda_{12}-\varepsilon\lambda_{21})$\\
 & $\varphi^{22}=(\lambda_{22}+\varepsilon\lambda_{21}u^2)\mathrm e^{\varepsilon(u^2)^2/2}$ && $\ne(0,0)$;\\
	&&&$\lambda_{11}+\lambda_{22}=1\bmod G_{1,1}^\sim$\\
	&&&$\lambda_{21}>0\ (\lambda_{21}\ne0),$\\
	&&&$\lambda_{12}\ge0\ (\lambda_{21}=0)\bmod G_{1,1}^\sim$\\

 \hline
5 & $\varphi^{11}=(\lambda_{11}-\varepsilon\lambda_{21}u^2)\mathrm e^{\varepsilon(u^2)^2/2}$,
 & $A^{\cap}_1,\ Q_\varepsilon,\ xQ_\varepsilon$
 & $\varepsilon=\pm1$,\\
 & $\varphi^{12}=(\varepsilon\lambda_{21}+\varepsilon\lambda_{11}u^2$ && $\lambda_{11}\lambda_{21}\ne0$,\\
 & $\qquad-\lambda_{21}(u^2)^2)\mathrm e^{\varepsilon(u^2)^2/2}$, && $\lambda_{22}=0$,\\
 & $\varphi^{21}=\lambda_{21}\mathrm e^{\varepsilon(u^2)^2/2}$, && $\lambda_{12}=\varepsilon\lambda_{21}$;\\
 & $\varphi^{22}=\varepsilon\lambda_{21}u^2\mathrm e^{\varepsilon(u^2)^2/2}$ &&\\
	&&&$\lambda_{11}=1,\quad\lambda_{21}>0$\\
	&&&$\bmod G_{1,1}^\sim$\\

 \hline

6 & $\varphi^{11}=\lambda_{11}|u^2|^m$, & $A^{\cap}_1,\ -m\p_{u^1}+u^2\p_{u^2}$
 & $\lambda_{22}\ne-\lambda_{11}$,\\
 & $\varphi^{12}=\lambda_{12}|u^2|^m/u^2$, && $\lambda_{11}\lambda_{22}\ne\lambda_{12}\lambda_{21}$,\\
 & $\varphi^{21}=\lambda_{21}u^2|u^2|^m$, && $(m,\lambda_{12},\lambda_{21})\ne(0,0,0)$,\\
 & $\varphi^{22}=\lambda_{22}|u^2|^m$ && $u^2\ne0$ if $(m,\lambda_{12})\ne(0,0)$;\\
	&&&$\lambda_{11}+\lambda_{22}=1\bmod G_{1,1}^\sim$\\

 \hline
\end{longtable}
\endgroup

\begin{remark}\label{rem:class1-canonical-exclusions}
For $n=0$, if $\varphi^{11}=0$ and $\varphi^{21}=\const\ne0$,
the system admits $x\p_{u^1}$ (case~5 of Table~\ref{tab2}).
If $\varphi^{11}=a\ne0$ and $\varphi^{21}=0$, it admits the ideal
$\{\beta(t,x)\p_{u^1}\mid\beta_t=a\beta_{xx}\}$
(case~6 of Table~\ref{tab2}).

For $n=1$, cases~4 and~5 of Table~\ref{tab3} admit
$Q_\varepsilon=-\varepsilon u^2\p_{u^1}+\p_{u^2}$,
$\varepsilon=\pm1$; case~5 also admits $xQ_\varepsilon$.
Both signs of $\varepsilon$ must be retained, since they are not
identified by real equivalence transformations.
In case~6 of Table~\ref{tab3}, if $m=\lambda_{12}=0$, we set
$\varphi^{12}=0$ also at $u^2=0$; the other entries of $\varPhi$
are then polynomial in $u^2$.
\end{remark}

\begin{proof}
We consider the cases $n=0$ and $n=1$ for class~\eqref{eq:I-class} separately.

\medskip\par\noindent
$\bm{n=0.}$
The first subsystem $\alpha^{21}\dot\varphi^{ab}=0$ of~\eqref{eq:I-class-define-system-n0}
splits the analysis into the cases where all entries of $\varPhi$ are constant and where at least one entry is nonconstant.

\medskip\par\noindent
$\bm{\dot\varphi^{ab}=0.}$
This case corresponds to the class of linear systems
\begin{gather}\label{eq:I-class-linear-system}
 U_t=\varLambda U_{xx},
\end{gather}
where $\varLambda$ is an arbitrary constant nonsingular $2\times2$ matrix
with $\tr\varLambda\ne0$.
Note that the matrix of the transformation of dependent variables
in~\eqref{eq:disc-transform-of-equivalence-Iclass-super}
is upper triangular. Therefore we must
use canonical forms under upper-triangular similarity,
rather than arbitrary similarity transformations from
$\mathrm{GL}(2,\mathbb R)$. We obtain four types:
\begin{gather*}
 \begin{pmatrix}1&0\\0&\lambda\end{pmatrix},\quad
 \begin{pmatrix}1&1\\0&1\end{pmatrix},\quad
 \begin{pmatrix}1&0\\0&1\end{pmatrix},\quad
 \begin{pmatrix}1&\eta\\1&1\end{pmatrix},
\end{gather*}
where the corresponding exclusions for $\lambda$ and $\eta$
are specified in cases~1--4 of Table~\ref{tab2}.
For the last type, $\eta\in\mathbb R\setminus\{1\}$ ensures $\det\varLambda\ne0$.
Applying Lie's algorithm to~\eqref{eq:I-class-linear-system}
for each of the above matrices $\varLambda$,
we obtain the maximal Lie invariance algebra $A^{\max}$ given in cases~1--4 of Table~\ref{tab2}.
Indeed, for any constant matrix $\varLambda$,
the algebra contains the operators $\beta^a(t,x)\p_{u^a}$,
where $\beta_t=\varLambda\beta_{xx}$, as well as Galilean
and projective symmetry operators. Direct substitution
into the determining system yields the basis
listed in the corresponding row of the table.
\medskip\par\noindent
$\bm{\dot\varphi^{ab}\ne0}$ for some $(a,b)$. Then $\alpha^{21}=0$.
We begin solving the remaining determining equations~\eqref{eq:I-class-define-system-n0} with the subsystem
		\begin{gather*}
	\left(\alpha^{22}u^2+\beta^2\right)\dot \varphi^{ab}+\left(\tau_t-2\xi_{x}\right)\varphi^{ab}+\alpha^{cb}\varphi^{ac}-\alpha^{ac}\varphi^{cb}=0.
\end{gather*}
Fixing the values of $t$ and $x$
in the same way as at the beginning of the proof of Theorem~\ref{thm:necessary-and-sufficient},
and taking $n=0$ into account,
we obtain the structural system for $\varphi^{ab}=\varphi^{ab}(u^2)$:
	\begin{gather}\label{eq:I-class-structure-system-n0}
		\left(k_{22}u^2+m_2\right)\dot \varphi^{ab}+k_{cb}\varphi^{ac}-k_{ac}\varphi^{cb}-m\varphi^{ab}=0,
\end{gather}
where $k_{21}=0$ because $\alpha^{21}=0$, and $m$ is the value of $2\xi_x-\tau_t$ at the fixed point $(t,x)$.
We split system~\eqref{eq:I-class-structure-system-n0} into two subsystems:
\begin{gather}\label{eq:characteristic-system-n0}
	\dfrac{\mathrm{d}u^2}{k_{22}u^2+m_2}=\dfrac{\mathrm{d}\varphi^{ab}}{m\varphi^{ab}+k_{ac}\varphi^{cb}-k_{cb}\varphi^{ac}}=\mathrm{d}p.
\end{gather}
We rewrite the first subsystem of~\eqref{eq:characteristic-system-n0} as
\begin{gather}\label{eq:I-class-first}
\dot u^2=k_{22}u^2+m_2,
\end{gather}
where
$\dot u^2:=\mathrm du^2/\mathrm dp$. Equation~\eqref{eq:I-class-first} serves as a classifier for splitting class~\eqref{eq:I-class} into inequivalent subclasses.

Transformations~\eqref{eq:disc-transform-of-equivalence-Iclass-super}
induce the equivalence group
\begin{gather}\label{eq:tr-u}
	\tilde u^2=l_2u^2+s_2
\end{gather}
of the class of equations~\eqref{eq:I-class-first},
where $l_2$ and $s_2$ are arbitrary constants with $l_2\ne0$.
The induced group splits this class into three orbits,
distinguished by 1) $k_{22}=m_2=0$, 2) $k_{22}=0$, $m_2\ne0$, and 3) $k_{22}\ne0$, $m_2=0$.
It suffices to consider one canonical representative in each orbit;
we therefore choose 1) $k_{22}=m_2=0$, 2) $k_{22}=0$, $m_2=1$, and 3) $k_{22}=1$, $m_2=0$.
We consider these cases separately.

\medskip\par\noindent
\textit{Case 1.} $k_{22}=m_2=0$. Combining the first and fourth equations of structural system~\eqref{eq:I-class-structure-system-n0} and using $\tr\varPhi\ne0$, we obtain $m=0$. Thus system~\eqref{eq:I-class-structure-system-n0} takes the form
\begin{gather}\label{eq:I-class-second-without-phi}
	k_{cb}\varphi^{ac}-k_{ac}\varphi^{cb}=0.
\end{gather}
Solutions of~\eqref{eq:I-class-second-without-phi} depend on $k_{11}$ and $k_{12}$. It suffices to consider the following subcases; their associated matrix families overlap in the scalar case:

1.1. $k_{12}=k_{11}=0$, \, $\varphi^{ab}$ are arbitrary smooth functions;

1.2. $k_{11}=0$, \, $k_{12}\ne0$, \, $\varphi^{21}=0$, \, $\varphi^{22}=\varphi^{11}$, \, $\varphi^{11}$ and $\varphi^{12}$ are arbitrary smooth functions;

1.3. $k_{11}\ne0$, \, $k_{12}=0$, \, $\varphi^{12}=\varphi^{21}=0$, \, $\varphi^{11}$ and $\varphi^{22}$ are arbitrary smooth functions.

In subcase 1.1, system~\eqref{eq:I-class} for $n=0$ takes the form
\begin{gather}\label{eq:I-class-system-n0-phi-arbitrary}
	\begin{aligned}
		&u^1_t=\p_x\left(\varphi^{11}(u^2)u^1_x+\varphi^{12}(u^2)u^2_x\right),\\
		&u^2_t=\p_x\left(\varphi^{21}(u^2)u^1_x+\varphi^{22}(u^2)u^2_x\right),
	\end{aligned}
\end{gather}
whose kernel of maximal Lie invariance algebras coincides with~\eqref{eq:I-class-kernel} for $n=0$.

In subcase 1.2, system~\eqref{eq:I-class} for $n=0$ is given by
\begin{gather}\label{eq:I-class-system-n0-phi1112-arbitrary}
	\begin{aligned}
		&u^1_t=\p_x\left(\varphi^{11}(u^2)u^1_x+\varphi^{12}(u^2)u^2_x\right),\\
		&u^2_t=\p_x\left(\varphi^{11}(u^2)u^2_x\right),
	\end{aligned}
\end{gather}
and its kernel of maximal Lie invariance algebras is given in case~15 of Table~\ref{tab2}.

In subcase 1.3, system~\eqref{eq:I-class} for $n=0$ takes the form
	\begin{gather}\label{eq:I-class-system-n0-arbitrary}
		\begin{aligned}
			&u^1_t=\p_x\left(\varphi^{11}(u^2)u^1_x\right),\\
			&u^2_t=\p_x\left(\varphi^{22}(u^2)u^2_x\right),
		\end{aligned}
	\end{gather}
and its kernel of maximal Lie invariance algebras is given in case~11 of Table~\ref{tab2}.

We now determine the conditions on $\varphi^{ab}(u^2)$ under which systems of classes~\eqref{eq:I-class-system-n0-phi-arbitrary},~\eqref{eq:I-class-system-n0-phi1112-arbitrary}, and~\eqref{eq:I-class-system-n0-arbitrary} admit extensions of the kernel of maximal Lie invariance algebras. For each class, we consider cases 2 and 3, beginning with class~\eqref{eq:I-class-system-n0-phi-arbitrary}.

If $k_{11}\ne0$ and $k_{12}\ne0$,
an upper-triangular equivalence transformation can
eliminate $k_{12}$ and reduce this subcase to 1.3.
The matrix families obtained in subcases 1.2 and 1.3 have the common scalar form
$\varPhi=f(u^2)I_2$, where $I_2$ is the $2\times2$ identity matrix. Its invariance algebra
contains both operators $u^1\p_{u^1}$ and
$u^2\p_{u^1}$; special functions $f$ will be examined
in the next two cases, leading to cases~7--10
of Table~\ref{tab2}.

\medskip\par\noindent
\textit{Case 2 for class~\eqref{eq:I-class-system-n0-phi-arbitrary}.} $k_{22}=0$, $m_2=1$.
Structural system~\eqref{eq:I-class-structure-system-n0} now takes the form
\begin{gather}\label{eq:I-class-structure-system-n0-2}
\begin{aligned}
 &\dot\varphi^{11}=m\varphi^{11}+k_{12}\varphi^{21},\qquad
 \dot\varphi^{12}=(m+k_{11})\varphi^{12}-k_{12}(\varphi^{11}-\varphi^{22}),\\
 &\dot\varphi^{21}=(m-k_{11})\varphi^{21},\qquad
 \dot\varphi^{22}=m\varphi^{22}-k_{12}\varphi^{21}.
\end{aligned}
\end{gather}
Solutions of this system depend on $k_{11}$ and
$k_{12}$.
The following subcases are possible.

\medskip\par\noindent
\textit{2.1.} $k_{11}=0$, $k_{12}=0$.
Direct integration gives
\begin{gather}\label{eq:c1-old-style-exp-common}
 \varPhi(u^2)=\mathrm e^{mu^2}\varLambda,
\end{gather}
where $\varLambda$ is a constant matrix.
For $m=0$, these are the linear systems considered above.
For $m\ne0$, we obtain exponential systems. A scalar matrix gives
case~9 of Table~\ref{tab2}; an upper-triangular matrix with
$\lambda_{21}=0$ and $\lambda_{22}\ne\lambda_{11}$ reduces to the
diagonal case~13 of Table~\ref{tab2}. If $\lambda_{21}=0$, $\lambda_{22}=\lambda_{11}$,
and $\lambda_{12}\ne0$, we obtain case~17 of Table~\ref{tab2} with $s=0$.
For $\lambda_{21}\ne0$, this branch is included in case~29 of Table~\ref{tab2} with $s=0$.

\medskip\par\noindent
\textit{2.2.} $k_{11}=0$, $k_{12}\ne0$.
Writing $a=k_{12}$ and integrating successively,
we obtain the following family
\begin{gather}\label{eq:I-class-structure-system-n0-2-gensol}
\begin{aligned}
 &\varphi^{11}=(\lambda_{11}+a\lambda_{21}u^2)\mathrm e^{mu^2},\\
 &\varphi^{12}=\bigl(\lambda_{12}+a(\lambda_{22}-\lambda_{11})u^2
                   -a^2\lambda_{21}(u^2)^2\bigr)\mathrm e^{mu^2},\\
 &\varphi^{21}=\lambda_{21}\mathrm e^{mu^2},\qquad
  \varphi^{22}=(\lambda_{22}-a\lambda_{21}u^2)\mathrm e^{mu^2}.
\end{aligned}
\end{gather}
For $m=0$, formulas~\eqref{eq:I-class-structure-system-n0-2-gensol} give a polynomial family. If $\lambda_{21}\ne0$, then
an equivalence normalization yields case~34 of Table~\ref{tab2}; if
$\lambda_{21}=0$ and $\lambda_{22}\ne\lambda_{11}$, then
we obtain case~25 of Table~\ref{tab2}. If $\lambda_{21}=0$ and
$\lambda_{22}=\lambda_{11}$, the matrix $\varPhi$ is constant and was considered above. For $m\ne0$, the relevant parameter conditions give case~31 of Table~\ref{tab2}, including $\lambda_{21}=0$, $\lambda_{22}\ne\lambda_{11}$. If $\lambda_{21}=0$ and $\lambda_{22}=\lambda_{11}$, then $\varPhi=\mathrm e^{mu^2}\varLambda$, which belongs to the previously considered branch~\eqref{eq:c1-old-style-exp-common}.

\medskip\par\noindent
\textit{2.3.} $k_{11}\ne0$.
Up to transformations~\eqref{eq:disc-transform-of-equivalence-Iclass-super},
we may set $k_{12}=0$. The general solution of
\eqref{eq:I-class-structure-system-n0-2} is then given by
\begin{gather*}
 \varphi^{11}=\lambda_{11}\mathrm e^{mu^2},\quad
 \varphi^{12}=\lambda_{12}\mathrm e^{(m+k_{11})u^2}, \quad
 \varphi^{21}=\lambda_{21}\mathrm e^{(m-k_{11})u^2}, \quad
 \varphi^{22}=\lambda_{22}\mathrm e^{mu^2}.
\end{gather*}
For $m=0$ and $\lambda_{21}\ne0$, we obtain case~32 of Table~\ref{tab2};
when $\lambda_{21}=0$, the nonconstant upper-triangular specializations
belong to cases~20 and~33 of Table~\ref{tab2}. For $m\ne0$, the general family is given in
case~29 of Table~\ref{tab2}, subject to its exclusions.
For $\lambda_{21}=0$ and $\lambda_{12}=0$, the diagonal exponential family with $m\ne0$ gives case~13 of Table~\ref{tab2} when $\lambda_{22}\ne\lambda_{11}$, and case~9 when $\lambda_{22}=\lambda_{11}$. If $\lambda_{12}\ne0$ and $\lambda_{22}=\lambda_{11}$, we obtain case~17 for $m\ne0$ and case~20 for $m=0$. The other specializations belong to cases~29 and~33, as follows from the remaining determining equations.

\medskip\par\noindent
\textit{Case 3 for class~\eqref{eq:I-class-system-n0-phi-arbitrary}.} $k_{22}=1$, $m_2=0$.
As in the preceding case, introduce
$\delta=k_{11}-1$. Then system
\eqref{eq:I-class-structure-system-n0} can be written as
\begin{gather}\label{eq:I-class-structure-system-n0-3}
\begin{aligned}
 &u^2\dot\varphi^{11}=m\varphi^{11}+k_{12}\varphi^{21},\\
 &u^2\dot\varphi^{12}=(m+\delta)\varphi^{12}
                      +k_{12}(\varphi^{22}-\varphi^{11}),\\
 &u^2\dot\varphi^{21}=(m-\delta)\varphi^{21},\qquad
 u^2\dot\varphi^{22}=m\varphi^{22}-k_{12}\varphi^{21}.
\end{aligned}
\end{gather}

\medskip\par\noindent
\textit{3.1.} $\delta=k_{12}=0$.
In this subcase, the general solution is
$\varPhi=(u^2)^m\varLambda$.
The case $m=0$ was considered above.
For $m\ne0$, a scalar matrix $\varLambda$ gives case~8 of Table~\ref{tab2}
(or case~10 for $m=-2$), and a diagonal matrix with distinct entries
gives case~12 of Table~\ref{tab2}. If $\lambda_{21}=0$, $\lambda_{22}=\lambda_{11}$,
and $\lambda_{12}\ne0$, the upper-triangular power family belongs
to case~16 (or case~22 for $m=-2$) of Table~\ref{tab2}, with $s=0$.
For $\lambda_{21}\ne0$, the systems belong to case~23 of Table~\ref{tab2} with $s=0$.

\medskip\par\noindent
\textit{3.2.} $\delta=0$, $k_{12}\ne0$.
Without loss of generality, write $a=k_{12}$.
Integrating system~\eqref{eq:I-class-structure-system-n0-3},
we obtain
\begin{gather}\label{eq:I-class-structure-system-n0-2-gensol3}
\begin{aligned}
 &\varphi^{11}=(\lambda_{11}+a\lambda_{21}\ln u^2)(u^2)^m,\\
 &\varphi^{12}=\bigl(\lambda_{12}+a(\lambda_{22}-\lambda_{11})\ln u^2
             -a^2\lambda_{21}\ln^2 u^2\bigr)(u^2)^m,\\
 &\varphi^{21}=\lambda_{21}(u^2)^m,\qquad
 \varphi^{22}=(\lambda_{22}-a\lambda_{21}\ln u^2)(u^2)^m.
\end{aligned}
\end{gather}
If $m\ne0$ and $(\lambda_{21},\lambda_{22}-\lambda_{11})\ne(0,0)$,
or if $m=0$ and $\lambda_{21}\ne0$, we obtain case~27 of Table~\ref{tab2}.
If $m=0$, $\lambda_{21}=0$, and $\lambda_{22}\ne\lambda_{11}$, we obtain case~28 of Table~\ref{tab2}, with a heat-equation ideal. If $m=0$, $\lambda_{21}=0$, and $\lambda_{22}=\lambda_{11}$, the matrix $\varPhi$ is constant and has already been considered. If $m\ne0$, $\lambda_{21}=0$, and $\lambda_{22}=\lambda_{11}$, the logarithmic terms vanish and we obtain the power specializations in cases~8, 10, 16, and~22 of Table~\ref{tab2}.

Admissible equivalence transformations are affine in the dependent variables; thus the logarithmic dependence in~\eqref{eq:I-class-structure-system-n0-2-gensol3} generally cannot be eliminated by these transformations and must be retained in the classification.

\medskip\par\noindent
\textit{3.3.} $\delta\ne0$.
Using transformation~\eqref{eq:disc-transform-of-equivalence-Iclass-super},
we may set $k_{12}=0$.
The general solution of~\eqref{eq:I-class-structure-system-n0-3}
then takes the form
\begin{gather*}
 \varphi^{11}=\lambda_{11}(u^2)^m,\quad
 \varphi^{12}=\lambda_{12}(u^2)^{m+\delta},\quad
 \varphi^{21}=\lambda_{21}(u^2)^{m-\delta},\quad
 \varphi^{22}=\lambda_{22}(u^2)^m.
\end{gather*}
Setting $s=-\delta$, we obtain the power family in case~23 of Table~\ref{tab2}.
If $\lambda_{21}=\lambda_{12}=0$, the matrix is diagonal or scalar:
for $m\ne0$ these are cases~8, 10, and~12 of Table~\ref{tab2}, while $m=0$ gives
constant matrices already considered. If $\lambda_{21}=0$ but
$\lambda_{12}\ne0$, equal diagonal entries give cases~16, 19,
and~22 of Table~\ref{tab2}, whereas distinct diagonal entries remain in case~23 when
$m\ne0$, and give cases~24 or~25 of Table~\ref{tab2} when $m=0$.
For $\lambda_{21}\ne0$, we obtain case~23, except for certain special cases. In particular, if $\lambda_{11}=0$, $\lambda_{21}\ne0$, and $\delta=m$, the first column of $\varPhi$ is constant. Checking the remaining determining equations then gives case~26 of Table~\ref{tab2}; the corresponding exponential case is listed as case~30 of Table~\ref{tab2}.

\textit{Cases 2 and 3 for classes~\eqref{eq:I-class-system-n0-phi1112-arbitrary} and~\eqref{eq:I-class-system-n0-arbitrary}.} We now turn to systems with $\varphi^{21}=0$.
In the subcase $\varphi^{11}=\varphi^{22}$, there is an operator
$u^2\p_{u^1}$ for arbitrary $\varphi^{11}$ and
$\varphi^{12}$. If $\varphi^{12}=0$, the operator
$u^1\p_{u^1}$ is also admitted, giving cases~7--10 of Table~\ref{tab2}. If $\varphi^{12}\ne0$, power and exponential
dependences yield the additional operators listed in
cases~16, 17, 19, and~20, with cases~15, 18, 21, and~22 of Table~\ref{tab2}
accounting for the other specializations.
If $\varphi^{11}\ne\varphi^{22}$, then
for a diagonal matrix only $u^1\p_{u^1}$ remains,
while for a constant first diagonal entry
the operators $\beta^1(t,x)\p_{u^1}$ are admitted,
with $\beta^1_t=\varphi^{11}\beta^1_{xx}$.
This gives cases~6, 11--14, and~24--25 of Table~\ref{tab2}.

We next consider matrices $\varPhi$ admitting more than one independent structural extension. In components, system~\eqref{eq:I-class-structure-system-n0} reads
\begin{gather}\label{eq:c1-trad-rank}
\begin{aligned}
 &(k_{22}u^2+m_2)\dot\varphi^{11}-m\varphi^{11}-k_{12}\varphi^{21}=0,\\
 &(k_{22}u^2+m_2)\dot\varphi^{12}-(m+k_{11}-k_{22})\varphi^{12}
       +k_{12}(\varphi^{11}-\varphi^{22})=0,\\
 &(k_{22}u^2+m_2)\dot\varphi^{21}-(m-k_{11}+k_{22})\varphi^{21}=0,\\
 &(k_{22}u^2+m_2)\dot\varphi^{22}-m\varphi^{22}+k_{12}\varphi^{21}=0.
\end{aligned}
\end{gather}
Such extensions correspond to linearly independent constant tuples
$(k_{22},m_2,m,k_{11}-k_{22},k_{12})$ satisfying~\eqref{eq:c1-trad-rank} for all $u^2$.
If two linearly independent structural generators have linearly dependent
projections onto the $u^2$-line, a nontrivial linear combination of them
has zero projection. These generators were already considered in
subcases~1.2--1.3, including their common scalar case
$\varPhi=f(u^2)I_2$, where $f$ need not be constant.
For independent projections, choose
$\p_{u^2}$ and $u^2\p_{u^2}$. Write $T=\tr\varPhi$ and
$g=\varphi^{21}$. The corresponding trace equations have the form
$\dot T=aT$, $u^2\dot T=bT$, and imply $\dot T=0$ since $T\ne0$.
Likewise $\dot g=cg$, $u^2\dot g=dg$, so $g$ is constant.
If $g\ne0$, the two structural equations give
$\dot\varphi^{11}=a_0g$ and $u^2\dot\varphi^{11}=b_0g$.
Since $g$ is a nonzero constant, $a_0=b_0=0$, and the remaining
components imply that $\varPhi$ is constant.
If $g=0$, the diagonal entries are constant, while
$f=\varphi^{12}$ satisfies $\dot f=af+b$ and $u^2\dot f=cf+d$.
Differentiating gives $\ddot f=a\dot f$ and $u^2\ddot f=(c-1)\dot f$;
therefore a nonconstant function $f$ is linear in $u^2$, $f=a_1u^2+b_1$ with $a_1\ne0$.
This is possible only when the diagonal entries of $\varPhi$ are distinct.
A shift of $u^2$ allows us to set $b_1=0$, yielding
case~25 of Table~\ref{tab2}.

We now use the remaining determining equations~\eqref{eq:I-class-define-system-n0} to check whether further Lie symmetry operators with coefficients depending on $x$ may occur.
Write $A=(\alpha^{ab})$, $b=(\beta^1,\beta^2)^{\top}$,
and $e_1=(1,0)^{\top}$, $e_2=(0,1)^{\top}$ for the standard basis vectors
of $\mathbb R^2$. From the equation involving $U_x$,
taking $\alpha^{21}=0$ into account, we obtain
\begin{gather}\label{eq:c1-trad-xgate}
 \xi_{xx}\varPhi+A_x\varPhi+\varPhi A_x
 +\dot\varPhi(A_xU+b_x)e_2^{\top}+\xi_tI_2=0.
\end{gather}
Differentiating the structural equation with respect to $x$ gives
\begin{gather}\label{eq:c1-trad-xdiff}
 (\alpha_x^{22}u^2+\beta_x^2)\dot\varPhi
   =2\xi_{xx}\varPhi+[A_x,\varPhi].
\end{gather}
The coefficient of $u^1$ in~\eqref{eq:c1-trad-xgate}
implies $\alpha_x^{11}\dot\varPhi e_1=0$.

We first consider the case where the first column
of $\varPhi$ is constant. For the operator
$X=\beta^1(t,x)\p_{u^1}$, the last two subsystems
of determining equations become
\begin{gather}\label{eq:c1-trad-heat}
 \beta_x^1\dot\varPhi e_1=0,\qquad
 \beta_t^1e_1=\beta_{xx}^1\varPhi e_1.
\end{gather}
If $\varPhi e_1=(\lambda_{11},0)^\top$ with
$\lambda_{11}\ne0$, then
$\beta_t^1=\lambda_{11}\beta_{xx}^1$. This yields
the symmetry ideals of vector fields $\beta^1(t,x)\p_{u^1}$
in cases~6, 14, 18--20, 24, 25, 28, and 33 of Table~\ref{tab2}.
If the second entry of the first column is nonzero,
then~\eqref{eq:c1-trad-heat} implies
$\beta_{xx}^1=\beta_t^1=0$, so the operator
$X=x\p_{u^1}$ appears, as in cases~5, 26, and~30 of Table~\ref{tab2}.

We next consider a first column depending on $u^2$.
If $\varphi^{21}=0$, write
$\varPhi=\bigl(\begin{smallmatrix}f&g\\0&h\end{smallmatrix}\bigr)$,
$A_x=\bigl(\begin{smallmatrix}a&\ell\\0&d\end{smallmatrix}\bigr)$,
and $b_x=(c,e)^\top$, with $\dot f\not\equiv0$.
The coefficient of $u^1$ and the first column in
\eqref{eq:c1-trad-xgate} give $a\dot f=0$ and
$(2a+\xi_{xx})f+\xi_t=0$; hence
$a=\xi_{xx}=\xi_t=0$. The first column of
\eqref{eq:c1-trad-xdiff} then gives $d=e=0$.
The remaining equations reduce to
$\ell(h-f)=0$ and $\ell(f+h+u^2\dot f)+c\dot f=0$.
If $\ell=0$, then $c=0$ as well, and no further Lie symmetry operators arise.
Otherwise $h=f$ and $(u^2+c/\ell)\dot f+2f=0$;
after subtracting the known generators, the additional operator
is proportional to $X=x(u^2+c/\ell)\p_{u^1}$. Renaming $c/\ell$ as $c$, it occurs only if
\begin{gather*}
 \varphi^{11}=\varphi^{22}=f(u^2),\qquad
 (u^2+c)\dot f+2f=0.
\end{gather*}
After shifting $u^2$, we obtain
$f(u^2)=\lambda_{11}(u^2)^{-2}$ and
$X=xu^2\p_{u^1}$, corresponding to cases~10, 21, and~22 of Table~\ref{tab2}.

If $\varphi^{21}\ne0$ and the first column is nonconstant,
from the first column of~\eqref{eq:c1-trad-xgate}
and~\eqref{eq:c1-trad-xdiff} we have
$\alpha_x^{11}=0$, $\xi_{xx}=-\alpha_x^{22}$.
When $\alpha_x^{22}=0$, the remaining conditions
yield no additional operators. Suppose that
$\alpha_x^{22}\ne0$ and put
$z=u^2+\beta_x^2/\alpha_x^{22}$.
Successive integration gives the matrix
\begin{gather*}
 \varPhi=\begin{pmatrix}
 gC/z&B-g^2C/z-ghC/z^2\\
 C/z&-gC/z-hC/z^2
 \end{pmatrix},\qquad C\ne0,
\end{gather*}
where $g,h,B,C$ are constants for fixed $t,x$.
Substituting the obtained expression for $\varPhi$ into~\eqref{eq:c1-trad-xdiff}
and eliminating the coefficients of $A_x$, we obtain $B=0$.
Thus $\det\varPhi=-BC/z=0$, contrary to the nondegeneracy assumption.
Thus all Lie symmetry operators with coefficients depending on $x$ have already been accounted for in this branch.
For a constant first column, equations~\eqref{eq:c1-trad-xgate}--\eqref{eq:c1-trad-xdiff} yield only the previously listed operators $\beta^1(t,x)\p_{u^1}$ and $x\p_{u^1}$ (in the respective cases).
Hence all such Lie symmetry operators have been accounted for when $n=0$.

Since cases~1--4 involve a constant matrix,
and all three possibilities for
$k_{22}u^2+m_2$ have been considered in the other cases, the resulting specializations together with
conditions~\eqref{eq:c1-trad-xgate}--\eqref{eq:c1-trad-heat}
yield cases~1--34 of Table~\ref{tab2}.
\medskip\par\noindent
$\bm{n=1}$. In this case, systems of class~\eqref{eq:I-class} have the form
\begin{gather*}
	U_t=\p_x\left[{\rm e}^{u^1}\varPhi\left(u^2\right)U_x\right].
\end{gather*}
The first subsystem $\alpha^{21}\dot\varphi^{ab}+\alpha^{11}\varphi^{ab}=0$
of determining system~\eqref{eq:I-class-define-system-n1}
leads to two cases: 1) $\alpha^{21}=\alpha^{11}=0$ and 2) $\alpha^{21}\ne0$. In the second branch, $\dot\varPhi=c\varPhi$ for some constant $c$; the admissible shift $\tilde u^1=u^1+cu^2$ reduces it to a constant matrix $\tilde\varPhi$.
Thus, in case 2 it suffices to study systems with a constant matrix of arbitrary elements.

\medskip\par\noindent
\textbf{1)} $\alpha^{21}=\alpha^{11}=0$.
The structural system corresponding to the second subsystem
of~\eqref{eq:I-class-define-system-n1} has the form
\begin{gather}\label{eq:I-class-structure-system-n1}
 (k_{22}u^2+m_2)\dot\varPhi
  +(k_{12}u^2-\mu)\varPhi+\varPhi K-K\varPhi=0,
 \quad K=\begin{pmatrix}0&k_{12}\\0&k_{22}\end{pmatrix},
\end{gather}
where $\mu$ is a constant determined by the corresponding
coefficients $\tau_t-2\xi_x+\beta^1$.
Up to transformations~\eqref{eq:tr-u}, there are again
three inequivalent subcases:
\begin{center}
1.1. $k_{22}=m_2=0$,\quad
1.2. $k_{22}=0$, $m_2=1$,\quad
1.3. $k_{22}=1$, $m_2=0$.
\end{center}
We consider them in turn.

\medskip\par\noindent
\textit{1.1.} $k_{22}=m_2=0$.
Taking the trace of~\eqref{eq:I-class-structure-system-n1}
and using $\tr\varPhi\ne0$, we obtain
$k_{12}=\mu=0$. Therefore there is no extension
of the kernel of maximal invariance algebras in this case.

\medskip\par\noindent
\textit{1.2.} $k_{22}=0$, $m_2=1$.
System~\eqref{eq:I-class-structure-system-n1}
takes the form
\begin{gather*}
\begin{aligned}
 &\dot\varphi^{11}=-(k_{12}u^2-\mu)\varphi^{11}+k_{12}\varphi^{21},\\
 &\dot\varphi^{12}=-(k_{12}u^2-\mu)\varphi^{12}
                 +k_{12}(\varphi^{22}-\varphi^{11}),\\
 &\dot\varphi^{21}=-(k_{12}u^2-\mu)\varphi^{21},\\
 &\dot\varphi^{22}=-(k_{12}u^2-\mu)\varphi^{22}-k_{12}\varphi^{21}.
\end{aligned}
\end{gather*}
Solving~\eqref{eq:I-class-structure-system-n1}, we obtain
\begin{gather*}
\begin{aligned}
 &\varphi^{11}=(\lambda_{11}+k_{12}\lambda_{21}u^2)
                  \mathrm e^{\mu u^2-k_{12}(u^2)^2/2},\\
 &\varphi^{12}=\bigl(\lambda_{12}+k_{12}(\lambda_{22}-\lambda_{11})u^2
                  -k_{12}^2\lambda_{21}(u^2)^2\bigr)
                  \mathrm e^{\mu u^2-k_{12}(u^2)^2/2},\\
 &\varphi^{21}=\lambda_{21}\mathrm e^{\mu u^2-k_{12}(u^2)^2/2},\\
 &\varphi^{22}=(\lambda_{22}-k_{12}\lambda_{21}u^2)
                  \mathrm e^{\mu u^2-k_{12}(u^2)^2/2}.
\end{aligned}
\end{gather*}
If $k_{12}=0$, the exponential factor
$\mathrm e^{\mu u^2}$ is removed by the transformation
$\tilde u^1=u^1+\mu u^2$; this gives the case
with a constant matrix.
If $k_{12}\ne0$, shifting $u^2$ eliminates
the linear term $\mu u^2$, and scaling
reduces $k_{12}$ to $-\varepsilon$,
$\varepsilon\in\{-1,1\}$.
We obtain the family in case~4 of Table~\ref{tab3}.
Both signs of $\varepsilon$ must be retained.

Substituting these functions into the remaining determining equations
shows that the additional operator
$\,xQ_\varepsilon$,
$Q_\varepsilon=-\varepsilon u^2\p_{u^1}+\p_{u^2}$,
is admitted if and only if
\begin{gather}\label{eq:c1-trad-n1-special}
 \lambda_{22}=0,\qquad
 \lambda_{12}=\varepsilon\lambda_{21}.
\end{gather}
These conditions give case~5 of Table~\ref{tab3}; otherwise,
we obtain case~4.

\medskip\par\noindent
\textit{1.3.} $k_{22}=1$, $m_2=0$.
In this case, we obtain the system
\begin{gather*}
\begin{aligned}
 &u^2\dot\varphi^{11}=-(k_{12}u^2-\mu)\varphi^{11}
                        +k_{12}\varphi^{21},\\
 &u^2\dot\varphi^{12}=-(k_{12}u^2-\mu+1)\varphi^{12}
                       +k_{12}(\varphi^{22}-\varphi^{11}),\\
 &u^2\dot\varphi^{21}=-(k_{12}u^2-\mu-1)\varphi^{21},\\
 &u^2\dot\varphi^{22}=-(k_{12}u^2-\mu)\varphi^{22}
                        -k_{12}\varphi^{21}.
\end{aligned}
\end{gather*}
The general solution of this system is given by
\begin{gather*}
\begin{aligned}
 &\varphi^{11}=(\lambda_{11}+k_{12}\lambda_{21}u^2)
                   (u^2)^\mu\mathrm e^{-k_{12}u^2},\\
 &\varphi^{12}=\bigl(\lambda_{12}+k_{12}(\lambda_{22}-\lambda_{11})u^2
                  -k_{12}^2\lambda_{21}(u^2)^2\bigr)
                   (u^2)^{\mu-1}\mathrm e^{-k_{12}u^2},\\
 &\varphi^{21}=\lambda_{21}(u^2)^{\mu+1}\mathrm e^{-k_{12}u^2},\\
 &\varphi^{22}=(\lambda_{22}-k_{12}\lambda_{21}u^2)
                   (u^2)^\mu\mathrm e^{-k_{12}u^2}.
\end{aligned}
\end{gather*}
Here these formulas are written for $u^2>0$; on $u^2<0$, the corresponding
real branches are obtained by replacing $(u^2)^\mu$ with $|u^2|^\mu$
and $(u^2)^{\mu\pm1}$ with $(u^2)^{\pm1}|u^2|^\mu$.
Using the transformation
$\tilde u^1=u^1-k_{12}u^2$ and the corresponding
similarity transformation of $\varPhi$, the factor
$\mathrm e^{-k_{12}u^2}$ is eliminated.
Thus the system reduces to the power family
\begin{gather*}
 \varphi^{11}=\lambda_{11}(u^2)^\mu,\quad
 \varphi^{12}=\lambda_{12}(u^2)^{\mu-1},\quad
 \varphi^{21}=\lambda_{21}(u^2)^{\mu+1},\quad
 \varphi^{22}=\lambda_{22}(u^2)^\mu,
\end{gather*}
given in case~6 of Table~\ref{tab3}. If the interval of definition contains
$u^2=0$, the trace equation for the transformed matrix is
$u^2\tr\dot\varPhi=\mu\tr\varPhi$.
Since $\tr\varPhi$ is nonzero at $u^2=0$, we have $\mu=0$.
Smoothness then requires $\lambda_{12}=0$ and gives the same
linear coefficient $\lambda_{21}$ on both sides of $u^2=0$.

\medskip\par\noindent
\textbf{2)} $\alpha^{21}\ne0$.
The first subsystem of~\eqref{eq:I-class-define-system-n1}
implies $\dot\varPhi=c\varPhi$,
where $c$ is a constant. As noted above, the
transformation $\tilde u^1=u^1+c u^2$ reduces
this branch to $\tilde\varPhi=\varLambda$,
that is, to the system
\begin{gather*}
 U_t=\varLambda\p_x\bigl(\mathrm e^{u^1}U_x\bigr).
\end{gather*}
Besides $\p_{u^2}$, additional linear generators of the form
$(a u^1+b u^2)\p_{u^2}$ are possible only if
\[
 [K,\varLambda]=0,\qquad K=\begin{pmatrix}0&0\\a&b\end{pmatrix}.
\]
If $\lambda_{12}\ne0$, this gives $a=b=0$; for
$\lambda_{12}=0$, the condition reduces to
$a(\lambda_{11}-\lambda_{22})+b\lambda_{21}=0$.
For a scalar matrix both parameters are free.
Thus we obtain cases~1--3 of Table~\ref{tab3}.

It remains to check whether the other determining
equations for $n=1$ admit further Lie symmetry operators
with coefficients depending on $x$. For this purpose,
we differentiate the structural system with respect to $x$
and substitute the result into the equation involving $U_x$.
In subcase~1.2 with $k_{12}\ne0$, splitting with respect to $u^2$
gives conditions~\eqref{eq:c1-trad-n1-special};
only when they hold does $xQ_\varepsilon$ appear.
In the power branch of subcase~1.3, the coefficients
$\xi_{xx}$ and $\alpha_x^{22}$ must satisfy
\begin{gather*}
\begin{aligned}
 &3\lambda_{11}\xi_{xx}+(\lambda_{12}-\mu\lambda_{11})\alpha_x^{22}=0,\\
 &(2\mu\lambda_{11}+\lambda_{12})\xi_{xx}
   +\mu(\lambda_{12}-\mu\lambda_{11})\alpha_x^{22}=0,\\
 &3\lambda_{21}\xi_{xx}
    +((1-\mu)\lambda_{21}+\lambda_{22})\alpha_x^{22}=0,\\
 &(2(\mu+1)\lambda_{21}+\lambda_{22})\xi_{xx}
    +(-\mu(\mu+1)\lambda_{21}+(\mu+2)\lambda_{22})\alpha_x^{22}=0.
\end{aligned}
\end{gather*}
The first two equations imply $\xi_{xx}=0$:
otherwise $\lambda_{11}=\lambda_{12}=0$,
contrary to $\det\varLambda\ne0$.
If $\alpha_x^{22}\ne0$, the remaining equations
again imply $\det\varLambda=0$.
Thus $\alpha_x^{22}=0$, and no new operators occur:
$\xi_{xx}=0$ means that $\xi$ is linear in $x$, whereas
$\alpha_x^{22}=0$ means that $\alpha^{22}$ is independent of $x$.
For a constant matrix, direct substitution into
\eqref{eq:I-class-define-system-n1}
also gives $\xi_{xx}=0$ and shows that no further Lie symmetry
operators with coefficients depending on $x$ occur.
Hence cases~1--6 of Table~\ref{tab3} exhaust
the extensions of $A_1^{\cap}$ for $n=1$.
\end{proof}

Thus, in the present paper we have found the equivalence group of the class of nonlinear diffusion systems, which enabled us to identify nine inequivalent subclasses whose systems admit extensions of the kernel of maximal Lie invariance algebras~\eqref{eq:kernel-of-mainsystem}. Equivalence transformations of the first identified class have been determined and its group classification has been carried out up to these transformations. The symmetry properties of diffusion systems from classes 2--9 will be studied in subsequent publications.

\vspace{3mm}
\textbf{Acknowledgments.} The authors are grateful to Professor R.~O.~Popovych for useful discussions, and to the participants of the seminar of the Department of Mathematical Physics and its head V.~M.~Boyko for discussing the results and providing valuable comments.

\textbf{Conflict of interest.} O.~O.~Vaneeva is a member of the Editorial Board of the Ukrainian Mathematical Journal. The handling editor of this paper was another member of the Editorial Board. The paper underwent independent peer review. O.~O.~Vaneeva was not involved in the review process or the decision on its publication.

\textbf{Funding.} This work was supported by grants from the Simons Foundation (1290607 and SFI-PD-Ukraine-00014586, O.~O.~V. and O.~A.~V.).

\textbf{Author contributions.} M.~I.~Serov and O.~O.~Vaneeva: formulation of the problem, general plan of the study, and verification of the results. O.~A.~Voloshyn: proofs of the main results.

  \label{literat}

\end{document}